\documentclass[aip,jcp,floatfix,reprint]{revtex4-1}
\usepackage[utf8]{inputenc}
\usepackage{lipsum}
\usepackage{float}
\usepackage{amsmath}
\usepackage{graphicx}
\usepackage{amssymb,amsmath}
\usepackage{geometry}
\usepackage{enumitem}
\usepackage{textcomp}
\usepackage{gensymb}
\usepackage{subcaption}
\usepackage{lipsum}
\usepackage{siunitx}
\usepackage[normalem]{ulem}
\usepackage{bm}
\usepackage{wrapfig}
\usepackage{listings}
\usepackage{placeins}
\usepackage[titletoc,title]{appendix}
\usepackage{pbox}
\usepackage{braket}
\usepackage{xcolor}
\usepackage{url}
\usepackage{setspace}
\usepackage{mathrsfs}
\usepackage{hyperref}
\usepackage{cleveref}
\usepackage{dsfont}
\usepackage{relsize}
\usepackage{tensor}
\newcommand{\CrD}{$\mathrm{Cr_2}$\ }
\newcommand{\FeP}{$\mathrm{Fe(P)}$\ }

\newcommand{\singlet}{${}^1 \mathrm{A}_{1g}$\ }
\newcommand{\quintet}{${}^5 \mathrm{A}_{1g}$\ }

\newcommand{\chimin}{\chi_\mathrm{min}}
\newcommand{\chimax}{\chi_\mathrm{max}}
\newcommand{\pmin}{p_\mathrm{min}}
\newcommand{\pmax}{p_\mathrm{max}}

\newcommand{\pkg}[1]{\textsc{#1}}

\newcommand{\abi}{\emph{ab initio}\ }

\graphicspath{ {figures/} }
\usepackage[nottoc]{tocbibind}

\newcommand{\remove}[1]{{}}
\newcommand{\add}[1]{{#1}}

\crefformat{equation}{(#2#1#3)}
\crefformat{figure}{#2#1#3}
\crefformat{section}{#2#1#3}

\definecolor{codegreen}{rgb}{0,0.6,0}
\definecolor{codegray}{rgb}{0.5,0.5,0.5}
\definecolor{codepurple}{rgb}{0.58,0,0.82}
\definecolor{backcolour}{rgb}{0.95,0.95,0.92}
\usepackage{xcolor}

\begin{document}

\title{Improved Stochastic Multireference Perturbation Theory for Correlated Systems with Large Active Spaces}

\author{James J. Halson}
\author{Robert J. Anderson}
\author{George H. Booth}
\email{george.booth@kcl.ac.uk}
\affiliation{Department of Physics, King's College London, Strand, London WC2R 2LS, U.K.}

\begin{abstract}
We identify the dominant computational cost within the recently introduced stochastic and internally contracted FCIQMC-NEVPT2 method for large active space sizes. This arises from the contribution to the four-body intermediates arising from low-excitation level sampled determinant pairs. We develop an effective way to mitigate this cost via an additional stochastic step within the sampling of the required NEVPT2 intermediates. We find this systematically improvable additional sampling can reduce simulation time by 80\% without introducing appreciable error. This saving is expected to increase for larger active spaces. We combine this enhanced sampling scheme with full stochastic orbital optimization for the first time, and apply it to find FCIQMC-NEVPT2 energies for spin states of an iron porphyrin system within (24,24) active spaces with relatively meagre computational resources. This active space size can now be considered as routine for NEVPT2 calculations of strongly correlated molecular systems within this improved stochastic methodology.
\end{abstract}

\maketitle

\section{Introduction}
\label{sect:Intro}

Full Configuration Interaction Quantum Monte Carlo (FCIQMC) is a computational method for performing efficient Full Configuration Interaction (FCI) calculations in a stochastic framework.
It has been found particularly effective for the \abi Hamiltonians of quantum chemistry in a basis of Hartree-Fock (HF) or other Self-Consistent-Field (SCF) orbitals \cite{booth/nature,doi:10.1063/1.3193710,Spencer_2012,doi:10.1063/1.3302277,Blunt/2015,PhysRevLett.114.033001,PhysRevLett.109.230201,PhysRevX.10.011041}.
Its fundamental operational assumption is that while the FCI wavefunction has a combinatorially large expansion in terms of N-electron basis states (Slater determinants), for real systems the FCI wavefunction is generally extremely sparse, with many determinants of extremely low wight \cite{BYTAUTAS200964}. 
FCIQMC realises the FCI wavefunction by stochastically applying a time-propagation matrix (derived from the sparse Hamiltonian of the system) onto a wavefunction that is sparsely represented by `walkers' -- automata that carry a signed weight representing FCI expansion coefficients. 
%It can be shown that in the limit of many applications of this propagator (or `timesteps' in the nomenclature of the FCIQMC algorithm) that the state converges on average to the ground state wavefunction of the system of interest. 
An efficient approximation designed to mitigate the sign problem introduces a small systematic error called the `initiator error' which can be rapidly reduced by increasing the number of walkers, systematically improving the resolution of the individual probability amplitudes of the states \cite{doi:10.1063/1.3302277,doi:10.1063/1.3624383}.

In the pursuit of improved accuracy and application to larger correlated systems of chemical interest, we have recently formulated FCIQMC within a stochastic and fully internally-contracted multireference perturbation theory framework \cite{doi:10.1063/1.5140086}.
In this approach, an active space of strongly correlated orbitals is first chosen, and subsequently sampled within the traditional FCIQMC dynamic.
After a steady-state convergence of the active space walker population is achieved, random sampling of effective higher-body operators {\em inside the active space} is performed \cite{doi:10.1063/1.4904313}.
These sampled quantities allow for an efficient post-processing computation of the perturbative contribution of the large number of degrees of freedom in the external space, as well as their coupling to the strongly correlated active space wave function.
This full internal contraction is essential to allow for the integrating out of the external space in the FCIQMC dynamic, and therefore avoidance of explicit sampling of excitations into this space, and the large increase in walkers (and decrease in sampling timestep) which this would incur \cite{doi:10.1063/1.4974177}.

Our initial work applied this approach to both 2nd-order Complete Active Space Perturbation Theory (CASPT2) and 2nd-order N-Electron Valence state Perturbation Theory (NEVPT2)\cite{doi:10.1063/1.5140086}.
However, it was shown that CASPT2 calculations based on FCIQMC observables are prone to numerical instability \cite{doi:10.1063/1.462209}. CASPT2 requires inverting the inherently noisy matrices estimated within the FCIQMC dynamic. This nonlinear operation amplifies any stochastic error in these quantities, and therefore the resulting CASPT2 correction is very sensitive to any noise in the FCIQMC wavefunction.
In contrast, strongly-contracted NEVPT2 avoids many of these difficulties of inverting stochastically derived matrices, as well as difficulties with intruder states \cite{C6SC03759C}.
Because of this, we found NEVPT2 to be stable and performed well, converging rapidly with number of walkers to the deterministic results, and allowing straightforward extension to active space sizes beyond those traditionally accessible.
Despite this success, for larger active spaces the sampling of high-rank tensors in the active space in FCIQMC-NEVPT2 can still become a bottleneck of the computation, and represents a significant fraction of the time on top of the original active space FCIQMC algorithm.
It is this challenge which we tackle in this work, developing and applying an improved sampling scheme for the FCIQMC-NEVPT2 method using both a stochastic CASCI, and also coupled with a self-consistent orbital optimisation obtained via a fully stochastic FCIQMC-CASSCF optimisation \cite{doi:10.1021/acs.jctc.5b00917}.

%, it is typical to perform a perturbation-theoretic correction to an initial calculation. 
%2nd Order Complete Active Space Perturbation Theory (CASPT2) is a common choice for this.
%It has been shown in some earlier work that CASPT2 is possibly ill-conditioned for FCIQMC applications, and that more reliable results might be obtained by instead applying 2nd Order N-Electron Valence State Perturbation Theory (NEVPT2)\cite{doi:10.1063/1.5140086}.
%More information on how this is performed is given in Section \ref{sect:FCIQMC_NEPT2} and in the reference.

The efficiency of FCIQMC is derived from a dual stochastisation of the problem -- both the wave function amplitudes are stochastically sampled, as well as the Hamiltonian (or propagator) governing the dynamics.
This is particularly efficient, as both quantities can be considered `sparse', and can be especially powerful if this sparsity can be effectively predicted in advance.
In the previous work on FCIQMC-NEVPT2, the sparsity in the sampling of higher-body active space quantities followed directly from the sparsity of the wave function amplitudes in terms of walkers.
However, in this work, we show that the exact and full accumulation of the higher-body quantity for any pair of stochastically sampled configurations can still lead to a significant and rapidly dominating overhead in computational cost in the stochastic NEVPT2.
We improve the FCIQMC-NEVPT2 algorithm to show that the introduction of an additional stochastic sampling in the accumulation of higher-body terms can exploit its sparsity and effectively reduce this cost, with minimal and systematically improvable loss of accuracy in the final results.
Furthermore, this approach fits well within the framework and motivation of the FCIQMC dynamic in terms of stochastically sampling sparse, high-dimensional quantities, especially where prior information can be used to enable this to be effective.

%However, NEVPT2 corrections represent a significant bottleneck in time over the base FCIQMC calculation due to their need to accumulate statistical representations of higher-rank tensors, which are combinatorial in system size, and so steps must therefore be taken to mitigate this computational cost.
%In this work we demonstrate some promising approaches that can reduce the time taken to perform the FCIQMC+NEVPT2 calculation within the Fortran FCIQMC package \pkg{NECI} (N-Electron Configuration Interaction solver). 

In Section \ref{sect:Cr2}, we show that this stochastic `importance sampling' on the accumulation of higher-body terms for a test $\mathrm{Cr_2}$ system results in a significant speedup even for a relatively small active space of 16 electrons and 16 orbitals, without significant loss of accuracy in the NEVPT2 energy compared to deterministic results.
We identify parameter regimes for this additional sampling that correspond to both low and high-fidelity realisations, and in all cases we obtain at least an order of magnitude speedup over an exact calculation.
% performed via the Python-based Simulations of Chemistry Framework (\pkg{PySCF}) C-accelerated Python package \cite{PYSCF}. 
Introducing more stringent thresholding parameters, can achieve close to another order of magnitude improvement, still with errors substantially less than one millihartree.

However, the magnitude of these efficiency improvements are expected to grow far larger as the active space increases in size.
To demonstrate this, in Sec.~\ref{sect:FeP} we apply the same approach to an iron porphyrin system, to investigate the performance for larger (24,24) active spaces, and consider the FCIQMC-NEVPT2 energetics of the low-lying spin dynamics. This is coupled with a stochastic self-consistent field optimisation of the active space to improve the reference state on which the NEVPT2 builds. We find rapid convergence of the NEVPT2 contributions with respect to number of walkers, with stochastic errors of order $\mathcal{O}[10^{-5}]$ with relatively small computational resources. We consider this size of active space as relatively straightforward within this enhanced sampling methodology for future multireference quantum chemical studies.

%in analogy with the work performed by Phung, Wouters and Pierloot, in which they investigate the energetics of this molecule using the Density Matrix Renormalisation Group augmented with CASPT2 (DMRG+CASPT2), and Coupled Cluster with Singles and Doubles and perturbative Triples (CCSD(T))\cite{PhungDMRG}. 
%We investigate some of the low-lying spin states of this molecule, which lie in different spatial symmetry sectors. 

\section{Computational challenges in FCIQMC-NEVPT2}

In the stochastic NEVPT2 algorithm, higher-body intermediate quantities are sampled within the walker algorithm in addition to the standard dynamics \cite{doi:10.1063/1.5140086}. Pairs of determinant amplitudes which are either three- or four-electron excitations of each other are stochastically sampled in an efficient manner, via spawnings between these determinants which sample, but do not change the wave function amplitudes. However, in larger active space sizes, we find the dominant cost in the stochastic NEVPT2 algorithm arises from the contributions to \emph{high-body intermediates} which are derived from pairs of determinants which are \emph{low-body excitations} of each other. In this section, we start with a short review of FCIQMC, and recap of the stochastic internally-contracted FCIQMC-NEVPT2 algorithm detailed for the first time in Ref.~\onlinecite{doi:10.1063/1.3193710}. We then identify the dominant computational bottleneck as arising from the `promotion' of low level excitation contributions to the four-body intermediates, as this requires several nested loops over common electron indices. Finally, a stochastic algorithm to effectively mitigate this cost is described in Sec.~\ref{sec:NEVPT2Bottleneck}.

\subsection{Review of FCIQMC}
\label{sect:FCIQMC_NEPT2}

Full configuration interaction quantum Monte Carlo is an algorithm for generating sparse stochastic realisations of the FCI wavefunction of a system described by an electronic Hamiltonian, $\hat{H}$.
Accurate expectation values that go beyond common single-reference approximations can then be obtained by averaging over their stochastic estimates.
We briefly summarise the features of FCIQMC that are important to the reader's understanding of the methodological developments in this work, and to define notation, with more complete technical descriptions to be found in Ref.~\onlinecite{booth_fciqmc_review}.
%Interested readers should follow references \cite{booth_fciqmc_review} and \cite{Spencer_2012} for a more complete description. 

The FCI wavefunction is expressed in a basis of $N$-electron Slater determinants, $\ket{D_\mathbf{i}}$, formed from a given set of underlying (in this work SCF molecular) orbitals (MOs), as
\begin{equation}
    \label{eq:CIexpansion}
    \ket{\Psi} = \sum_\mathbf{i} C_\mathbf{i} \ket{D_\mathbf{i}},
\end{equation} 
where the determinant index $\mathbf{i}$ represents the specific MO occupation vector defining the determinant.
In this basis of Slater determinants, the Hamiltonian between configurations, $H_{\mathbf{ij}}$, is sparse, derived from the nature of the Hamiltonian as a two-body operator.
We write a discrete snapshot of the wave function amplitudes in terms of walkers at a given timestep, $\tau$, as $\mathbf{C}(\tau)$.
Iteratively and stochastically applying the linearised imaginary-time propagation converges $\mathbf{C}(\tau)$ to a discrete representation of the ground-state wave function in FCIQMC, as
\begin{equation}
    \label{eq:discreteTimeLinearImagSchrodinger}
    \mathbf{C}(\tau + \Delta \tau) = [\mathds{1} - \Delta \tau (\mathbf{H}-\mathds{1}S)] \mathbf{C}(\tau),
\end{equation}
where $\Delta \tau$ is a small discrete time-step, and $S$ is a variable energy `shift', introduced to enable a population control of the number of walkers. The key concept of FCIQMC is that for many choices of system and underlying MO basis, the vast majority of configurational amplitudes are very small, and thus treated as $C_{\mathbf{i}}=0$ in the discrete snapshots at any single given time.
Therefore at each iteration, the walker population only stores the (signed) non-zero amplitudes as a sparse, yet dynamically evolving population, in an efficient and distributed memory fashion.
This leads to an efficient compression of the FCI state without an explicit exponential memory footprint.

An important part of the algorithm for this work is the `spawning' step \cite{doi:10.1021/acs.jctc.5b01170}, whereby $n_s$ excitations ($\mathbf{j}$) are stochastically generated from each occupied determinant ($\mathbf{i}$), and new walker weight is created on them as
\begin{equation}
    \Delta C_\mathbf{j} ^\text{(spawn)}=-C_\mathbf{i}\Delta\tau\frac{H_\mathbf{ij}}{n_s P_\text{gen}(\mathbf{i}|\mathbf{j})} .
\end{equation}
These newly-created walkers are then subject to annihilation with walkers of opposite sign, and initiator criteria which aims to limit the growth of sign-incoherent amplitudes.
The initiator criteria introduces a systematic error into the average wave function, which can be systematically reduced via an increase in walker number.
Ensuring convergence of properties calculated from FCIQMC with respect to the number of walkers is an important test of the method, and this rate of convergence is key for the efficiency of the approach for a given system.

Observables computable from the FCI wavefunction, such as the energy, spin eigenvalues, and reduced density matrices can be extracted from the FCIQMC representation by a statistical averaging process \cite{doi:10.1063/1.4762445,doi:10.1063/1.4904313}.
For scalar observables the memory footprint of accumulating these statistics is of little consequence, however as we shall see for the higher-order tensor observables required for perturbation theory, the efficient distributed storage of these quantities requires consideration \cite{doi:10.1063/1.4986963}.

\subsection{Stochastic Reduced Density Matrix Accumulation}
\label{sect:NECI_RDMs}
The normal-ordered reduced density matrix of rank $N$ (The $N$-RDM) of an arbitrary state $\ket{\Psi}$ is defined as
\begin{equation}
\label{eq:NRDM_general}
    \Gamma_{i_1 i_2 \dots i_N,j_1 j_2 \dots j_N}^{(N)} = \bra{\Psi} \hat{a}^\dagger_{i_1} \hat{a}^\dagger_{i_2} \dots  \hat{a}^\dagger_{i_N}
    \hat{a}_{j_1} \hat{a}_{j_2}  \dots  \hat{a}_{j_N}\ket{\Psi}
\end{equation}
Which in an orthogonal CI basis described by Eq. \eqref{eq:CIexpansion} reduces to
\begin{equation}
\label{eq:NRDM_FCI}
\begin{split}
    \Gamma_{i_1 i_2 \dots i_N,j_1 j_2 \dots j_N}^{(N)} = & \sum_{\mathbf{ij}} \\
    C_\mathbf{i}^*C_\mathbf{j}  \bra{D_\mathbf{i}} & \hat{a}^\dagger_{i_1} \hat{a}^\dagger_{i_2} \dots  \hat{a}^\dagger_{i_N}
    \hat{a}_{j_1} \hat{a}_{j_2}  \dots  \hat{a}_{j_N} \ket{D_\mathbf{j}}
\end{split}
\end{equation}
where the factor $\bra{D_\mathbf{i}} \hat{a}^\dagger_{i_1} \hat{a}^\dagger_{i_2} \dots  \hat{a}^\dagger_{i_N}
    \hat{a}_{j_1} \hat{a}_{j_2}  \dots  \hat{a}_{j_N} \ket{D_\mathbf{j}}$ is equivalent to a sign depending on the parity of ordering of the Fermionic operators.% \cite{SOURCEreference on NEVPT2?}.
We may therefore calculate elements of the reduced density matrix using the stochastic CI coefficients $\set{C_\mathbf{i}}$ provided by FCIQMC.
In practice to eliminate bias from the bilinearity of the RDM with respect to the wavefunction coefficients, $C_\mathbf{i}$ and $C_\mathbf{j}$ are drawn from two different ``replicas'' of the FCIQMC wavefunction which are specified by identical parameters, but which use different seeds for random number generation, allowing their evolution by the stochastic application of Eq. \eqref{eq:discreteTimeLinearImagSchrodinger} to generate statistically independent distributions of determinant weights \cite{doi:10.1063/1.4904313,doi:10.1063/1.4986963}.

In the FCIQMC propagation, occupied determinants generate spawning events between connected determinants.
These connections may be leveraged to also serve as choices of $\ket{D_\mathbf{i}}$ and $\ket{D_\mathbf{j}}$ in Eq. \eqref{eq:NRDM_FCI}.
The excitation rank of the connection is the number of individual electron replacements required to transform the occupation vector $\mathbf{i}$ into $\mathbf{j}$, or equivalently, the number of excitation operators ${\hat E}_{ij} = \hat{a}_i^\dagger \hat{a}_j$ with $i \neq j$ that are required to transform one determinant into the other.
It is easy to see from Eq. \eqref{eq:NRDM_FCI} that if $\ket{D_\mathbf{i}}$ and $\ket{D_\mathbf{j}}$ are separated by an excitation rank greater than $N$, then their contribution to the $N$-RDM will vanish.
When the excitation rank is equal to $N$, the excitation $\ket{D_\mathbf{i}} \rightarrow \ket{D_\mathbf{j}}$ describes uniquely a single element of the $N$-RDM.
For example, considering the second-rank RDM $\Gamma^{(2)}$, if $\ket{D_\mathbf{j}}$ is a double excitation of $\ket{D_\mathbf{i}}$, such that $\ket{D_\mathbf{j}} =  \hat{a}_i^\dagger \hat{a}_j^\dagger \hat{a}_k \hat{a}_l \ket{D_\mathbf{i}}$, then this corresponds to only the element $\Gamma^{(2)}_{ij,kl}$ of the 2-RDM.

When the excitation is of rank less than $N$, the determinant pair will appear in all elements of the $N$-RDM containing the excitation indices, and any other combination of repeated indices occupied in both $\ket{D_\mathbf{i}}$ and $\ket{D_\mathbf{j}}$.
If $\ket{D_\mathbf{j}}$ is a single excitation of $\ket{D_\mathbf{i}}$, so $\ket{D_\mathbf{j}} =  \hat{a}_i^\dagger \hat{a}_j \ket{D_\mathbf{i}}$, then this corresponds to the unique element $\Gamma^{(1)}_{i,j}$ of the 1-RDM, but also all elements of the 2-RDM of the form $\Gamma^{(2)}_{ik,jk}$, where $k$ is any of the other occupied indices of $\ket{D_\mathbf{i}}$ or $\ket{D_\mathbf{j}}$, which are necessarily the same as they differ in only one index by construction.

Therefore, for all generated connections $\ket{D_\mathbf{i}} \rightarrow \ket{D_\mathbf{j}}$, it is possible to `promote' them to a set of off-diagonal elements of higher-body RDMs. The number of such elements grows combinatorially as the difference between the excitation rank and the RDM rank increases (for example, to promote the single excitation to a 3-RDM contribution requires constructing all unique unordered \emph{pairs} of shared indices). It is this cost of promoting low excitation rank pairs of determinants to high-rank quantities (RDMs and other partially-transformed intermediates required for the stochastic NEVPT2) which we focus on in this section.

\subsection{Stochastic NEVPT2}
Quantitative accuracy of quantum chemical calculations is routinely achieved by applying a Second-order perturbation-theoretic (PT2) correction to self-consistently obtained complete active space wave functions \cite{doi:10.1063/1.462209,FINLEY1998299,ANGELI2001297}.
Applying the second-order excitation operators directly to this state gives rise to an `internally-contracted' formalism, where the coupling of the active space to the external space is described by high-order correlation functions (RDMs) in the active space. The avoidance of explicitly computing excitations from the active space into the external space dramatically decreases the scaling of the approach for large, chemically-relevant external spaces. Complete Active Space Perturbation Theory (CASPT2) is a popular choice in this spirit, as it is computationally the cheapest approach \cite{doi:10.1021/j100377a012, doi:10.1063/1.462209}.
However, N-electron Valence State Perturbation Theory (NEVPT2) represents an alternative to CASPT2 that, while strictly more computationally expensive, is size-consistent and also does not suffer from the adverse influence of `intruder states' \cite{C6SC03759C}. More importantly within our stochastic formulation, it has been shown to be more stable and tolerant of random errors, due to its avoidance of inversion of stochastically-derived matrices, with almost all of the energy contributions linear with respect to these quantities \cite{PhysRevB.98.085118}.

The perturbers of NEVPT2 are divided into disjoint classes according to the redistribution of electrons they cause between the core, active, and virtual orbital spaces \cite{doi:10.1063/1.1361246, ANGELI2001297, doi:10.1063/1.1515317}.
The spaces are denoted $S_l^{(k)}$, where $l$ is a composite index signifying the external space indices involved in the perturbation, and $k$ is the net displacement of electrons into the active space.
The application of the perturbation operators $\hat{V}^{(k)}_l$ to the correlated active space wavefunction as a whole (as opposed to the individual determinants in the CI expansion), is key to the internal contraction.

The formulation of NEVPT2 is based on Dyall's Hamiltonian, defined by
\begin{equation}
	\hat{H}^D = \hat{H}_i + \hat{H}_v
\end{equation}
\begin{equation}
	\hat{H}_i \equiv 
		\sum_{i\in\text{core}} \epsilon_i \hat{E}_{ii}+
		\sum_{r\in\text{virtual}} \epsilon_r \hat{E}_{rr} + C
\end{equation}
\begin{equation}
    \hat{H}_v \equiv \sum_{ab\in\text{CAS}} h_{ab}^{\text{eff}} \hat{E}_{ab} +\frac{1}{2} \sum_{abcd\in\text{CAS}} \langle ab | cd \rangle \left(\hat{E}_{ac}\hat{E}_{bd} - \delta_{bc}\hat{E}_{ad}\right)
\end{equation}
where $\epsilon_i$ and $\epsilon_r$ are the external orbital energies obtained from the diagonalisation of the generalised Fock matrix
\begin{equation}\label{eq:general_fock_matrix}
F_{pq} = h_{pq}+\sum_{rs} \overline{\Gamma_{rs}} \left[(pq|rs)-\frac{1}{2}(pr|qs)\right]
\end{equation}
where all indices extend over the entire orbital space, and hence the external space orbital energies depend on the correlated active space density matrix.

The energies associated with the second order perturbative correction due to each perturber are given by
\begin{align}
	\label{eq:E2_slk_delta}
	\mathcal{E}_l^{(k)} &= -N_l^{(k)}\left(
		\frac{\langle\Psi^{(0)}|\hat{V}^{(k)}_l{}^\dagger[\hat{H}^v, \hat{V}^{(k)}_l]|\Psi^{(0)}\rangle}{N^{(k)}_l} + \Delta^{(k)}_l
	\right)^{-1}  \\
	&\equiv
	-N_l^{(k)}\left(\frac{h^{(k)}_l}{N^{(k)}_l} + \Delta^{(k)}_l\right)^{-1}
\end{align}
where $N_l^{(k)}\equiv \langle\Psi^{(0)}|\hat{V}^{(k)}_l{}^\dagger\hat{V}^{(k)}_l|\Psi^{(0)}\rangle$ are the perturber normalisations, \add{and $ \Delta^{(k)}_l$ are the matrix elements of the external part of Dyall's Hamiltonian (which amount to sums and differences of orbital energies)}.
By these expressions the sc-NEVPT2 energies can be evaluated given the matrix elements of the Hamiltonian, and the active space density matrices up to rank four.

A recent development by some of the authors has allowed for the required active space expectation value tensors to be stochastically computed within the FCIQMC active space dynamics, allowing for large active space sparse sampling within this framework \cite{doi:10.1063/1.5140086}.
In this work, the stochastic RDM sampling capabilities were adapted for the estimation of three-body RDMs. Specifically, since the four-body RDM only appears in the NEVPT2 equations contracted with other quantities, the FCIQMC algorithm need only estimate the a contracted three-body active-space intermediate derived from the sampling of the four-body RDM, given by
\begin{equation}
\overline{G^{1}}_{a'b'c',a b c} =  \sum_{def}^{\text{CAS}}\sum_{\mu\nu\sigma\tau}
    \braket{de|fa}
    \Gamma^{(4)}_{c_\sigma' b_\tau' d_\mu e_\nu, c_\nu f_\nu b_\tau a_\sigma'}   \label{eqn:4RDMContract}
\end{equation}
where the Latin subscripts refer to spatial orbitals and the Greek subscripts denote spin indices \cite{doi:10.1063/1.5140086}.
It is therefore sufficient to contract the FCIQMC-estimated $\Gamma^{(4)}$ elements on-the-fly and store only elements of the contracted quantity $\overline{G^1}$, which obviates the explicit storage of $\Gamma^{(4)}$.

This on-the-fly contraction for sampled four-body excitations scales linearly with the size of the system, through the explicit enumeration required over \add{$a$, the only orbital index in the definition of $\overline{G^{1}}$ which is not determined by the excitation}.
However, the \emph{promotion} of \mbox{(semi-)stochastic} excitations of ranks one, two, and three (by consideration of common occupied indices in the pair of determinants) into $\Gamma^{(4)}$ elements in Eq.~\ref{eqn:4RDMContract} is the new dominant cost. This is because it entails the enumeration of all combinations of occupied spin orbitals in common between the bra and ket determinants. For instance, for a sampling of a single electron excitation within the active space, an $\mathcal{O}[N^3]$ operation is required to compute all the contributions to $\Gamma^{(4)}$, followed by another $\mathcal{O}[M]$ operation to perform the contraction on each, where $N$ is the number of active space electrons, and $M$ is the number of active space orbitals. For double electron excitations, there are only $\mathcal{O}[N^2]$ contributions, however, this is offset against the larger overall number (and frequency of sampling) of double excitations in the active space. Finally, for three-electron active space excitations, there are a linear number of `promoted' contributions required to consider. It is the cost of this promotion which we aim to reduce via an additional stochastic sampling strategy within the accumulation of the NEVPT2 intermediate of Eq.~\ref{eqn:4RDMContract}.

\subsection{Tackling the NEVPT2 Bottleneck}
\label{sec:NEVPT2Bottleneck}

Calculating $\overline{G^1}$ based on the stochastic generation scheme outlined in Section \ref{sect:NECI_RDMs} means that the cost of promoting generated (especially low-body) excitations to elements of the 4-RDM is dominant in comparison to any other stage of the calculation, and will increase as the number of electrons in the active space increases. 
%For each triple excitation that is generated, a linear-scaling loop is required, for each double excitation, a quadratic loop, and for each single excitation, a cubic-scaling loop is needed.
%This promotion procedure is the bottleneck of FCIQMC-NEVPT2, and so we now present a new strategy to specifically mitigate the cost of this part of the calculation.
The basic assumption of the adaptation we present is that many of the sampled $C_\mathbf{i}^*C_\mathbf{j}$ products are small in magnitude, and that it is therefore practical and efficient to randomly select a subset of the corresponding promotion combinations rather than explicitly enumerating all of them.
The importance of each contribution is determined by a monotonic function of $C_\mathbf{i}^*C_\mathbf{j}$, and this function is therefore used to determine the number of randomly selected promoted contributions to the 4-RDM term in Eq.~\ref{eqn:4RDMContract}. This therefore efficiently exploits the sparsity in the sampling of the promoted 4-RDM contributions to the stochastic NEVPT2 energy in such a way that is systematically improvable by increased sampling time, or modifying the parameters defining the sampling function.

\begin{figure}
    \centering
    \includegraphics[width=\columnwidth]{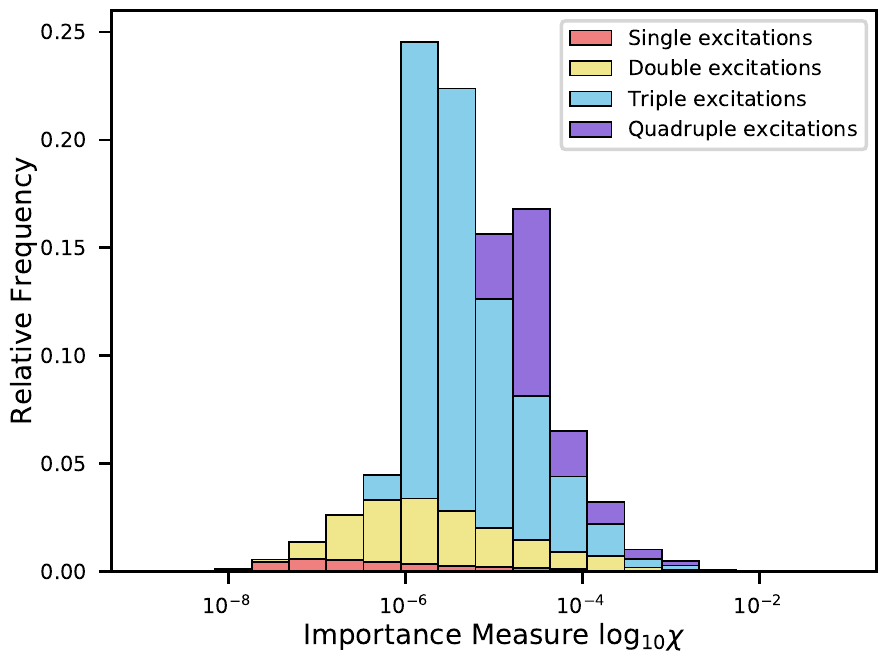}
    \caption{Histogram of magnitude of normalised $C_\mathbf{i}^\ast C_\mathbf{j}$, generated during an FCIQMC calculation for a chromium dimer ($\text{Cr}_2$) system at equilibrium geometry for a CAS of 16 electrons in 16 active orbitals, using a cc-pVTZ basis. The total number of walkers present was 100,000. The contributions are separated into the different excitation levels sampled for each determinant connection $\mathbf{i} \rightarrow \mathbf{j}$. As can be seen, the magnitudes of the connections are concentrated between $10^{-8}$ and $10^{-3}$.}
    \label{fig:chi_example}
\end{figure}

We define the function
\begin{equation}
    \chi_\mathbf{ij} = \frac{C_\mathbf{i}^\ast C_\mathbf{j}}{\left| \braket{\Psi | \Psi} \right|}
\end{equation}
as an appropriately normalised measure of the importance of a given excitation to the 4-RDM.
%, with the number of elements of the 4-RDM which the pair of determinants contributes to depending on the excitation level between them, and given by the binomial coefficient, $n_\mathrm{comb}$.
A representative plot of the distribution of $\chi_\mathbf{ij}$ for sampled determinant pairs in the (16,16) CAS of the Chromium dimer in a cc-pVTZ basis is shown in Fig.~\ref{fig:chi_example}. This illustrates that the magnitudes of the sampled contributions to the RDMs vary over six orders of magnitude. Whilst most contributions are $\mathcal{O}[10^{-6}]$, this is substantially less than the largest contributions, and therefore admitting only a sparse sampling of this distribution over this range seems reasonable \footnote{In Fig.~\ref{fig:chi_example}, triple and quadruple excitations are sampled frequently relative to singles and doubles. These higher two excitation levels are sampled purely for the purposes of higher-body RDM sampling and do not affect the dynamics. Therefore, their relative ratio of sampling compared to single and double excitations is not a physical quantity, but rather a user-defined choice. This is controlled by a `granularity' parameter (detailed in previous work \cite{doi:10.1063/1.5140086}), which in this example was set to one, ensuring significant sampling of the higher-body excitations. In production simulations, we often choose to sample the higher-rank excitations less frequently. In this case the promotion of the more expensive single and double excitations increasingly dominate the cost, and the issue of mitigating their undesirable scaling becomes even more important.}. Furthermore, determinant pairs differing by a single excitation are amongst the smallest contributions in terms of individual magnitude, however are required to contribute to the high-rank $\Gamma^{(4)}$ $\mathcal{O}[N^3]$ times per sampled pair. It is clear that computational efficiency can be improved by a sparse sampling of these promoted contributions to high-rank objects, based on their magnitude of $\chi_{\bf ij}$.

Two thresholds on this value are defined: $\chimin$ and $\chimax$. If $\chi_\mathbf{ij} < \chimin$ then the promotion loops are abandoned completely, and for $\chi_\mathbf{ij} > \chimax$, then every possible
element of the 4-RDM which the pair of determinants contributes to depending on the excitation level between them is computed explicitly and deterministically. We refer to this total number of elements by the binomial coefficient $n_\mathrm{comb}$.
For determinant pairs with $\chi_\mathbf{ij}$ between $\chimin$ and $\chimax$, there is defined a ``promotion probability'' $p(\chi)$ that interpolates between $\pmin=p(\chimin)$ and $\pmax=p(\chimax)$. It is in this region that the stochastic promotion occurs. 

In this case, instead of the true number of promotion combinations to $\Gamma^{(4)}$, given by $n_\mathrm{comb}$ which depends on the excitation rank, we instead choose a random subset of size $p(\chi)\times n_\mathrm{comb}$ to promote. 
This additional level of stochastisation is done by precomputing and storing the $n_\mathrm{comb}$ possible index combinations in the promotion for a given excitation level at initialisation. During the RDM sampling for an active space excitation, the promotion of the contribution is then found via calculation of $p(\chi)$, and then repeatedly drawing random elements from this enumeration, until a total of $p\times n_\mathrm{comb}$ have been drawn. The weighting of each contribution to $\Gamma^{(4)}$ is then increased by a factor of $\frac{1}{p(\chi)}$ to unbias for this stochastic selection, before contraction on-the-fly down to $\overline{G^1}$. \add{It should be noted that while the estimation of $\chi_{\bf ij}$ in a stochastic simulation may be slightly biased due to the non-linear biases intrinsic to the sampling, the fact that this is unbiased by the same factor ensures that no error is introduced due to any potential discrepancy compared to the exact value of $\chi_{\bf ij}$.}

\begin{center}
\begin{figure}[t]
%    \centering
    \includegraphics[width=\columnwidth ]{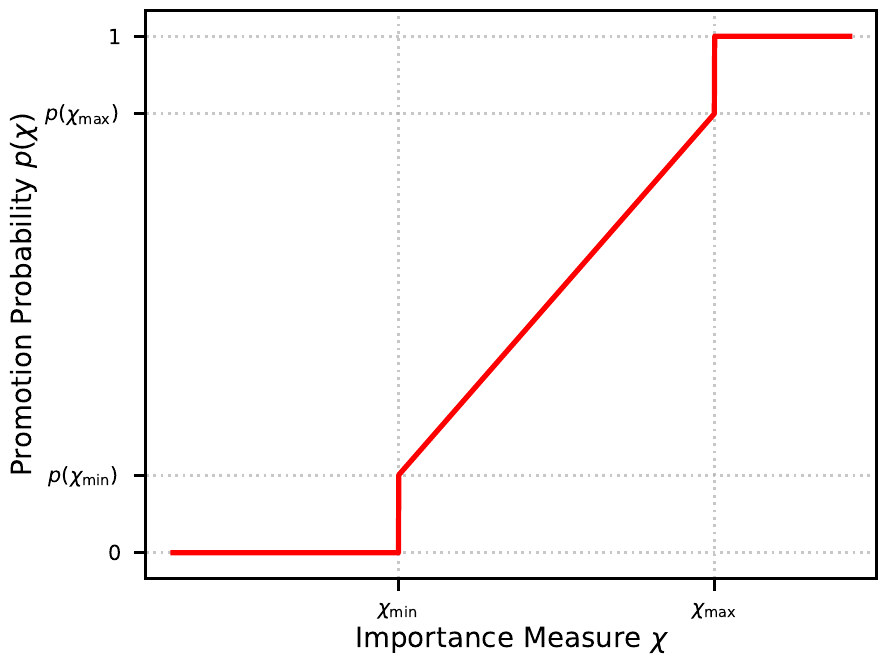}
    \caption{Illustrative example of the probability function $p(\chi)$ used to perform stochastic promotion events with a fidelity proportional to their normalised contribution to a single RDM element, $\chi$. Below $\chimin$, no promotion is performed, and above $\chimax$, all contributions are included deterministically. In between, a fraction of the total number of combinations, $p(\chi)$, is chosen that interpolates between a minimum and maximum value, to reduce the computational work in calculating these contributions while retaining accuracy of the final accumulation of the NEVPT2 intermediate.}
    \label{fig:ramp}
\end{figure}
\end{center}

In practice we found a probability distribution for $p(\chi)$ consisting of a ``linear ramp'' (illustrated in Fig.~\cref{fig:ramp}) is effective and simple to interpret. This is described by the equation
\begin{equation}
    p(\chi) = \begin{cases}
     \hfil 0&, \: \chi < \chimin\\
      \hfil 1&, \: \chi > \chimax\\
    \pmin + \frac{\pmax - \pmin}{\chimax - \chimin} (\chi - \chimin),& \text{otherwise}
    
    \end{cases},
\end{equation}
where $p(\chi)=0$ and $p(\chi)=1$ are respectively understood to mean that the promotion loops are either aborted, or that the deterministic enumeration is used instead. The deterministic enumeration is distinct from actually performing the stochastic enumeration until $n_\text{comb}$ combinations have been selected, as in the stochastic promotion, it is possible to draw the same combination more than once.
By inclusion of a lower limit for $\chimin$, this form for the promotion probability also ensures that $p(\chimin)$ can be set to a non-zero value. This is important to reduce noise from infrequent stochastic contributions from very small RDM contributions, as otherwise these would be renormalised against a vanishingly small probability, which can introduce a large variance into the sampling. 
At some larger promotion probability, the loss of sequential data access and the cost of drawing random numbers combine to render the stochastisation ineffective due to its intrinsic overhead compared to explicit enumeration. Therefore $p(\chimax)$ can be set below unity to avoid the introduction of additional computational overhead in cases where the stochastic enumeration would require a time investment comparable to simply calculating the contribution deterministically.

Also attempted was a logarithmic ramp, where $p(\chi)$ is proportional to $\log_{10}(\chi)$. This was found to provide very little timing improvement as it did not in fact assign low probabilities to almost all of the terms in the sum, and so was roughly equivalent to a step-function form of $p(\chi)$ with a step at $\chimin$. Other more sophisticated forms of of $p(\chi)$ could be an avenue for improvement, possibly subject to some iterative optimisation procedure. It should be noted that this scheme is formally unbiased in its sampling of $\Gamma^{(4)}$ and $\overline{G^1}$ for the FCIQMC sampled wave function, assuming that all contributions where $\chi_{\bf ij} < \chimin$ can be safely neglected. This ensures that the deterministic promotion result is converged to upon increased sampling iterations. The replica scheme which is used in the sampling of all RDMs is also used in the computation of $\chi_{\bf ij}$, determining the frequency of contributions to the sampled intermediates. This ensures that all quantities depend linearly on stochastic variables. Finally, we also note that while in this work, the stochastic promotion scheme is only applied to the 4-RDM contributions of the $\overline{G^1}$ intermediate in Eq.~\ref{eqn:4RDMContract}, in principle, the same scheme could be applied to lower-rank objects if required, such as the 3-RDM. However, as we will see, the 4-RDM promotion consists of the dominant contribution to the computational effort of a calculation, and therefore the added benefits from this would be small.

\subsection{Computational Procedure}
\label{sec:CompProc}
Given a sufficient number of walkers to sample the active space wave function, we are capable of resolving $\overline{G^{1}}$ with improvable accuracy in the limit of a large number of sampling iterations.
Estimation of $\overline{G^{1}}$ is the dominant cause of both computational expense and statistical error, and so every effort is made to ensure that all other relevant quantities are sampled to high accuracy before such a costly process is initiated.
The walker number is first grown to a target population at which dynamic shift variation is activated.
The population then continues to evolve under initiator-FCIQMC dynamics until certain key metrics (e.g. energy estimators) are seen to stabilise about equilibrium values. 
Then, the determinants with the highest walker occupation are chosen to make up the `deterministic space' of the semi-stochastic adaptation, which entails the exact propagation of Hamiltonian connections both within and coupled to this chosen space of highly occupied determinants\cite{PhysRevLett.109.230201, doi:10.1063/1.4920975}.
This measure reduces fluctuations in the largest walker occupations, and consequently lowers random error in expectation values.
Soon after the initialisation of semi-stochastic propagation, RDM estimation can begin.

The use of Dyall's Hamiltonian in NEVPT2 requires that the molecular orbitals are semi-canonical with respect to the generalised Fock matrix \add{defined in Eq. \ref{eq:general_fock_matrix}}. 
A canonicalising transformation must therefore be applied to the external orbital space prior to the evaluation of the NEVPT2 subspace energies---an operation highly sensitive to the accuracy of the one-body RDM due to its non-linear dependence on the random errors.
As is clear from Eq.\ref{eq:E2_slk_delta}, the subspace energies are also sensitive to error in the normalisation of each perturber.
The norms and all subspace energies with the exceptions of $\mathcal{E}_{i}^{(+1)}$ and $\mathcal{E}_{r}^{(-1)}$ are calculable from $\Gamma_{(3)}$, thus it is important to prioritise the accurate resolution of the lower-rank RDMs before incurring the dominant computational expense of estimating the $\Gamma^{(4)}$-dependent NEVPT2 intermediates.
Once both the walker population is equilibrated 
%(save for a small error due to the initiator approximation) about the exact solution, 
and the cheaper RDMs are well resolved, the $\overline{G^{1}}$ estimate is accumulated  using the stochastic sampling scheme detailed above. This is performed over multiple statistically independent realisations to ensure that faithful error estimates can be achieved, and until the errors in $\mathcal{E}_{i}^{(+1)}$ and $\mathcal{E}_{r}^{(-1)}$ have been reduced within acceptable bounds. All NEVPT2 computation from the sampled intermediates, as well as the CASSCF iterations and RHF calculations are performed via a modified version of the \pkg{PySCF} code\cite{pyscf}, while the FCIQMC sampling is developed for this purpose within the \pkg{NECI} code\cite{booth_fciqmc_review}.

\section{Benchmarking the approach: The Chromium dimer}
\label{sect:Cr2}
The impact of stochastic promotion on the efficacy of FCIQMC-NEVPT2 can be rigorously benchmarked by applying the method to an active space for a real molecular system, where we do not exceed the limit of deterministic FCI solvers \add{to allow comparison to exact, noiseless results}.
For this purpose, a 16 electron, 16 spatial orbital CAS was chosen from the RHF solution for Cr${}_2$ in a cc-pVTZ basis at an equilibrium bond length of \SI{1.6788}{\angstrom}
%\add{, as well as at a stretched geometry of \SI{2.4}{\angstrom}, where the wave function is significantly multiconfigurational}
. Cr${}_2$ is known to be highly multireference \cite{PhysRevResearch.2.012015,doi:10.1063/1.4906829} and is a system for which FCIQMC has previously been shown to be effective \cite{booth_fciqmc_review}.
The FCIQMC-NEVPT2 calculations for \CrD were performed using stochastic promotion probability ramps with $\chimin$ at $10^{-9}$, $10^{-8}$ and $10^{-7}$, and $\chimax$ at each integer order of magnitude between $10^{-9}$ and $10^{-3}$.
In each case, the minimum and maximum stochastic promotion probabilities were left at their default values of 0 and 1 respectively. 
Each FCIQMC-NEVPT2 calculation was repeated with five different random number generation seeds, to provide an estimate of the standard error on the NEVPT2 energies obtained. These are compared against the result of an exact CASCI-NEVPT2 calculation. %calculation performed in \pkg{PySCF} 
%with a result of $E_\mathrm{CASCI}=\SI{-1.558333}{\hartree}$.
%which required a total of 3348 core hours on a workstation PC.

\begin{figure}[t]
    \centering
    \includegraphics[width=\columnwidth]{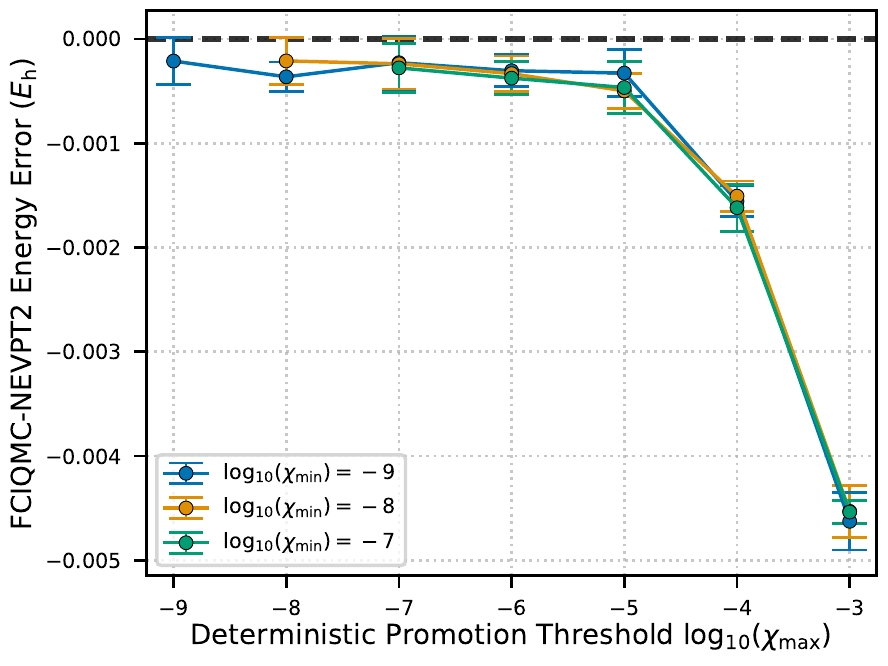}
    \caption{FCIQMC-NEVPT2 energy error as a function of stochastic promotion cutoffs, $\chimin$ (varied between $10^{-9}-10^{-7}$) and $\chimax$ (varied between $\chimin$ and $10^{-3}$). The error is relatively independent on this range of $\chimin$, but exhibits a rapid increase for large $\chi_\mathrm{max}$. Errors are substantially below \SI{1}{\milli \hartree} up to $\chimax = 10^{-5}$ for all series, which demonstrates small errors are possible at a reduced computational cost. Remaining systematic errors in the deterministic promotion limit are due to the inadequate sampling of the lower-rank RDMs which define the sc-NEVPT2 perturber normalisations, and would be further reduced with additional sampling time or walker number.
    }    \label{fig:CrD_energy}
\end{figure}

The effect of this stochastic promotion sampling inside the FCIQMC main loop on the final sampled FCIQMC-NEVPT2 energy \add{at the equilibrium geometry} can be seen in Fig.~\ref{fig:CrD_energy}. It is clear that values of $\chimin$ below $10^{-7}$ do not introduce any additional systematic or stochastic errors into the final result, and these contributions can be safely neglected. However, setting the value of $\chimax$ in the promotion sampling to greater than $10^{-5}$ is found to rapidly increase the systematic error in the final result. The fact that this wider and coarser sampling manifests as a systematic rather than random error is due to the non-linearities in the NEVPT2 expressions involving $\Gamma^{(4)}$ in the $\mathcal{E}_r^{(-1)}$ and $\mathcal{E}_i^{(+1)}$ subspaces, as given in Eq.~\ref{eq:E2_slk_delta}. The fact that the converged energies where $\chimax$ is less than $10^{-5}$ still show a small systematic error of $\mathcal{O}[10^{-4}]E_{\mathrm h}$ is indicative of the random errors in quantities due to the original walker sampling of the FCIQMC, rather than the promotion sampling, which are also non-linearly transformed in the NEVPT2 expression (primarily in the 1-RDM in the semi-canonicalisation of the orbitals) \cite{PhysRevB.98.085118,PhysRevLett.115.050603}. This remaining systematic error is able to be reduced via increased sampling time of the appropriate intermediates, as well as increasing walker number, all of which will decrease further the random errors in the sampled RDMs, and hence the systematic error in the NEVPT2 energies. Because of this, we adjusted the simulation procedure to ensure that the lower-rank RDMs are highly sampled to minimise this residual error, before the more costly accumulation via promotion sampling of $\Gamma^{(4)}$ is initiated, as detailed in Sec.~\ref{sec:CompProc}. However, it is clear that the promotion sampling scheme does not affect the accuracy of the final FCIQMC-NEVPT2 energy for $\chimax<10^{-5}$ and $\chimin<10^{-7}$\remove{, which are likely to continue to be reasonable choices of these parameters for $\mathcal{O}[10^{-4}]E_{\mathrm h}$ accuracy for a wide range of systems}.
\add{To give an initial indication of the transferability of these parameters to more strongly correlated systems and to ensure that no further systematic error is introduced, they were also applied to the chromium dimer at a stretched bond length of \SI{2.4}{\angstrom}, where the system is significantly more multiconfigurational. All parameters save for the number of walkers, which must necessarily be increased for a more multiconfigurational system to ensure the initiator error is controlled, were kept the same. This calculation also resulted in systematic error to the deterministic calculation substantially below a millihartree ($0.25(5){\mathrm m}E_{\mathrm h}$ compared to exact results). We therefore assert that the parameter set $\chimax<10^{-5}$ and $\chimin<10^{-7}$ will continue to be a reasonable choice for a range of further systems.}

%results of this comparison are shown in Figure \ref{fig:CrD_energy}, with the timing data (broken down by the amount of time taken by each level of excitation as described in section \cref{sect:NECI_RDMs}) is shown in \cref{fig:CrD_timing}. 
%It should be noted that all of these calculations were performed with a granularity of 1, and so in principle the proportion of the total calculation time consumed by the stochastic promotion would be even greater for a higher granularity that would be typical in a practical calculation. 
%As evidenced by the plot, increasing the minimum cutoff and the maximum cutoff can result in significant speed up of the calculation.

\begin{figure}[ht]
    \centering
    \includegraphics[width=\columnwidth]{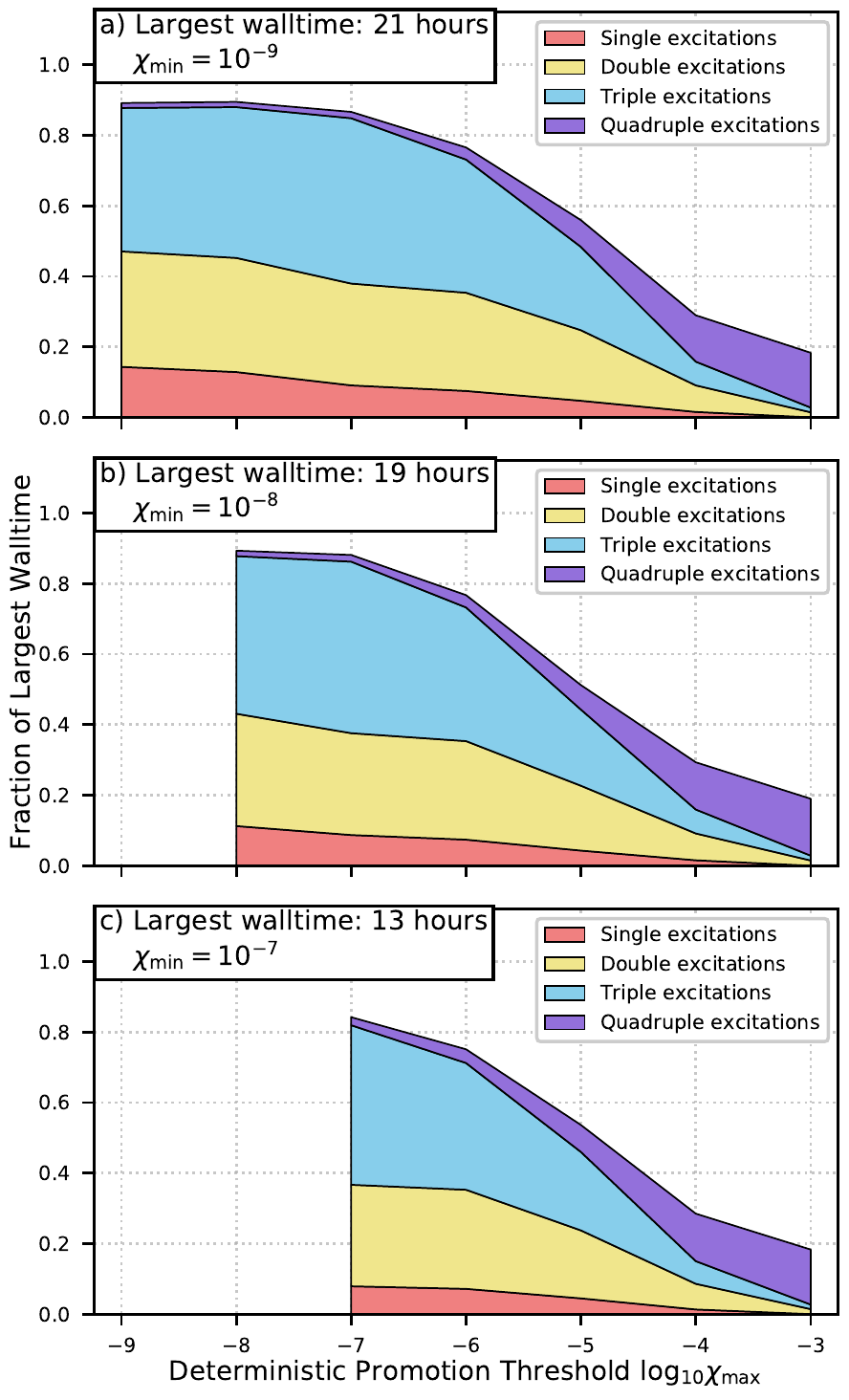}
    \caption{Time taken for contracted 4-body (${\overline{G^1}}$) accumulation in FCIQMC-NEVPT2 for various promotion probability ramp parameters. Time is measured proportional to the time for a $\chimin=\chimax$ calculation in each plot, where no stochastic promotion is performed. a) $\chimin=10^{-9}$, b) $\chimin=10^{-8}$, c) $\chimin=10^{-7}$. Large proportions of time are taken in promotion when there is little stochastic promotion sampling (small $\chi$ parameters) which is rapidly reduced as these parameters are increased. Total times refer to the $\chimin=\chimax$ calculation in each plot (the longest calculation, without stochastic promotion).}
    \label{fig:Crd_timing}
\end{figure}

    The benefits of this stochastic promotion sampling are evident in Fig.~\ref{fig:Crd_timing}, where the reduction in computational time is shown for the same ranges of $\chimin$ and $\chimax$ parameters. With the most restrictive parameters of $\chimin=\chimax=10^{-9}$, where essentially all but the most insignificant promotion events are deterministically enumerated, it can be seen that approximately 90\% of the calculation time is spent in enumerating the promoted contributions to $\overline{G^1}$ via samples from $\Gamma^{(4)}$. This is further broken down into the different excitation levels of the determinant pairs, where the fraction of time promoting single excitations is still significant, despite their relatively small number of sampled contributions compared to double and triple excitations. As $\chimax$ is increased, the fraction of the total walltime spent sampling these promotions substantially reduces. Furthermore, this decrease is most substantial for the lower-rank excitations, eventually leaving the sampled contribution of the most numerous quadruple excitations as the dominant expense. This improvement in the fraction of time spent in the promotion step is reproduced for all three plots, where $\chimin$ varies between $10^{-7}$ and $10^{-9}$. \remove{The $\chimin$ and $\chimax$ parameters identified in Fig.~\ref{fig:CrD_energy} as appropriate to avoid introducing any additional systematic error into the final FCIQMC-NEVPT2 energy leads to a reduction in total walltime by approximately a factor of four.} 
    
    \add{Comparing the timing of the least restrictive parameters in Figure \ref{fig:Crd_timing}a ($\chimin=\chimax=10^{-9}$)  with the $\chimin=10^{-7}$, $\chimax=10^{-5}$ point in Figure \ref{fig:Crd_timing}c (the parameters that were shown in Figure \ref{fig:CrD_energy} to comfortably continue to give sub-millihartree errors), we can see that the timing is improved by around a factor of four without loss of accuracy.}
    This factor is expected to increase for larger active space sizes, where the number of promotions increases binomially with the number of active electrons. 
    %\add{However, it is also important to check that there is transferability of these values to avoid introduction of unwanted systematic error for systems with a even more multiconfigurational character. To test this, we stretch the Chromium bond to 1.5 times equilibrium geometry, where the FCIQMC-NEVPT2 energy can measured with the same $\chimin=10^{-7}$ and $\chimax=10^{-5}$ parameters in the stochastic promotion sampling. The number of walkers was doubled in this active space, to 10 million compared to the equilibrium geometry, which is standard when requiring sampling of a more multiconfigurational state. However, all other parameters were kept the same, and since the $\chi$ values scale with the norm of the wave function The error compared to the deterministic result}
    \add{Having verified the approach and a parameter set to use with it, }
    we therefore turn to a final application in a larger active space, where we also include a full stochastic orbital optimisation.

\section{Stochastic CASSCF-NEVPT2 applied to the spin states of Iron Porphyrin}
\label{sect:FeP}
Neutral iron(II) porphyrin, denoted \FeP, where P is $(\mathrm{C}_{20}\mathrm{N}_{4}\mathrm{H}_{12})^{2-}$, is a biologically-important complex from which haem is derived. As such a key molecule, it is unsurprising that much other modern multireference quantum chemistry has been applied to it, including a DMRG-CASPT2 study by Phung \emph{et al.} \cite{doi:10.1021/acs.jctc.6b00714}, as well as previous FCIQMC-CASSCF studies \cite{doi:10.1021/acs.jctc.5b01190,doi:10.1021/acs.jpca.7b12710,doi:10.1021/acs.jctc.8b01277, Weser2020} amongst other theoretical approaches\cite{doi:10.1021/ct200597h,doi:10.1021/ct500103h, doi:10.1021/acs.jctc.9b01255}. Part of the difficulty of calculations involving this molecule (and the origin of its biological importance) is the many closely-separated low-energy spin states of the system. Theoretical studies, especially those based on CASPT2, often find an incorrect quintet ground state\cite{doi:10.1021/ct200597h,doi:10.1021/acs.jctc.8b01277} attributed to a poor PT2 description of the semi-core correlation \cite{doi:10.1021/acs.jctc.6b00714,doi:10.1021/acs.jctc.8b01277}, while experimental and other higher-accuracy computational approaches point to a triplet ground state\cite{doi:10.1021/ja00300a021,doi:10.1021/ct500103h,doi:10.1021/acs.jctc.8b01277}.

In this work, we focus on the lesser-studied excited \singlet and \quintet spin states, which we consider with a stochastic FCIQMC-CASSCF orbital optimisation and subsequent FCIQMC-NEVPT2 with the detailed stochastic promotion. This is performed within a large (24,24) active space including a double d-shell on the iron and substantial semi-core correlation in the CAS, beyond conventional means to solve. The basis is chosen to be cc-pVTZ on the chemically important iron atom, and cc-pVDZ on all other atoms, correlating all electrons. This basis is likely too small for true quantitative accuracy, as well as the neglect of any relativistic effects, but will still represent a stern test for the methodology. This large active space also reduces the sensitivity of the final results to the choice of initial active space orbitals, as well as the reference Hartree--Fock state, for which there are a number of low-energy competing solutions. Geometries used for this study were taken from Ref. \onlinecite{doi:10.1021/acs.jctc.6b00714} within a $D_{4h}$ symmetry, where they were optimised at the PBE0/def2-TZVP level separately for each spin state.

\subsection{FCIQMC-CASSCF Orbital Optimisation}

\begin{figure}[t]
    \centering
    \includegraphics[width=\columnwidth]{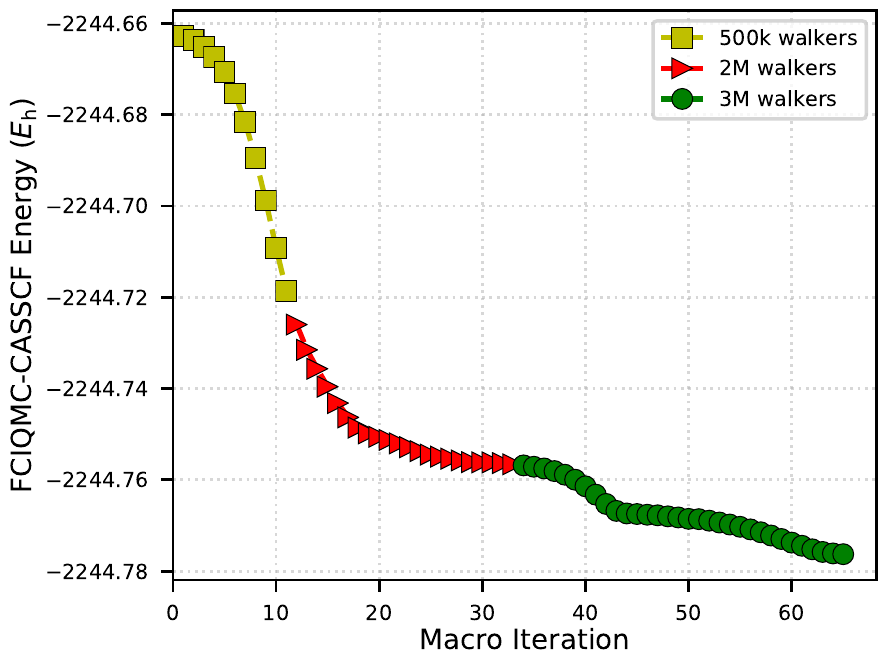}
    \caption{FCIQMC-CASSCF energy convergence as the (24,24) active space orbitals are optimised for the \singlet state of Iron (II) Porphyrin. The quintet is not shown but demonstrated similar convergence. Walker numbers were varied during the course of the convergence to minimise overall cost for the calculation, where earlier iterations are performed with a coarser active space description.}
    \label{fig:CASSCF}
\end{figure}

The FCIQMC-CASSCF procedure is detailed in Ref. \onlinecite{doi:10.1021/acs.jctc.5b00917}, and we largely follow this here.
The CAS orbitals for the iron porphyrin calculations are selected based on an initial restricted Hartree-Fock calculation performed with \pkg{PySCF}.
The 24 orbitals for the CAS (12 occupied and 12 unoccupied) were selected via a meta-L{\" o}wdin orbital analysis of the RHF/ROHF molecular orbitals, ensuring that the orbitals with the largest weight on the N $2p$ and $3p$, and Fe $3d$ and $4d$ atomic orbitals were selected to comprise the initial active space. The active space for each spin state was optimised independently, where the FCIQMC dynamic in the active space was used to sample the 2-RDM used for subsequent orbital optimisation in a two-step procedure \cite{SUN2017291}.

% \begin{figure}[t]
% \centering
%   \includegraphics[width=\columnwidth]{1A1g_orbs.png}
%   \caption{CASSCF active space orbital plots for FeP \singlet\label{fig:1A1g_orb_plots}}
% \end{figure}

% \begin{figure}[t]
% \centering
%   \includegraphics[width=\columnwidth]{5A1g_orbs.png}
%   \caption{CASSCF active space orbital plots for FeP \quintet\label{fig:5A1g_orb_plots}}
% \end{figure}

Figure~\ref{fig:CASSCF} shows the convergence of the FCIQMC-CASSCF energy with respect to orbital optimisation iterations for the \singlet state, with the \quintet state showing similar convergence. This convergence is artificially slow, since we start the optimisation with a small number of FCIQMC walkers at each iteration for a coarse description of the CAS wave function. This leads to a comparatively large error in the subsequent orbital optimisation, but these iterations are particularly cheap to perform. As convergence slows, the number of FCIQMC walkers is increased to improve the accuracy of the final iterations, up to 3 million walkers, at which point we find that the FCIQMC-CASSCF is converged for this system. This is still a comparatively small number of walkers for FCIQMC, and this full optimisation can be considered relatively routine. \add{Plots of the active space orbitals obtained from this procedure are shown in the supporting information for this article\cite{Knizia2013,Knizia2015}.}

%After the selection of the initial CAS from the \pkg{PySCF} orbitals, an interative CASSCF calculation was performed whereby \pkg{NECI} was used to estimate a 2-body RDM, from which an orbital rotation could be generated to a new set of orbitals, which may then be put back into \pkg{NECI}, and the process repeated. This converges towards a set of natural orbitals in which the energy $E_\text{CASSCF}$ are as low as possible. Throughout the convergence, parameters of the NECI calculation (notably the walker number, and the number of iterations for which RDM data were accumulated) were adjusted on-the-fly to strike a balance between quality of RDM estimation, and quick iterations. Once the energy was judged to be stable with iterations, the procedure was halted and the final orbitals used for the subsequent NEVPT2. The CASSCF convergence curves are shown in Figure \ref{fig:CASSCF}

\subsection{FCIQMC-NEVPT2 Energies}

\begin{figure}[t]
\centering
   \includegraphics[width=\columnwidth]{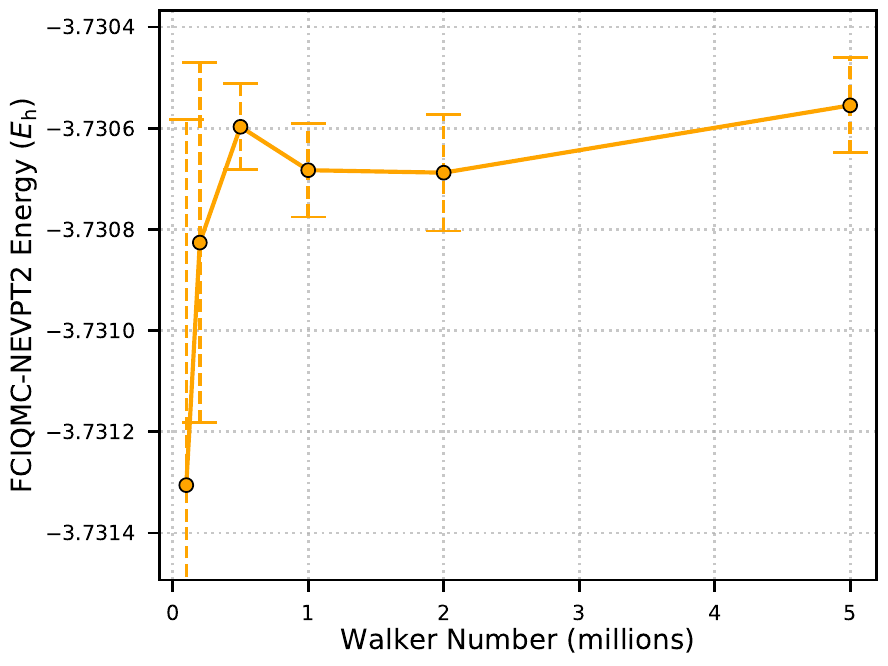}
   \caption{Convergence of the FCIQMC-NEVPT2 energy for the \singlet state of iron(II) porphyrin in a (24,24) active space of FCIQMC-CASSCF optimised orbitals. Convergence is shown as the number of active space walkers is increased. Stochastic errors are estimated from five independent seeds.}
   \label{fig:1A1g_nevpt2} 
%\caption{Convergence of FCIQMC-NEVPT2 energy with respect to \TBD{parameter} for iron porphyrin system.}
\end{figure}

The stochastic NEVPT2 was performed on top of the optimised orbitals, with the simulation procedure outlined in Sec.~\ref{sec:CompProc}. The convergence of the stochastic NEVPT2 energy with respect to number of walkers sampling the active space is given in Fig.~\ref{fig:1A1g_nevpt2} for the \singlet state. The initial phase of the calculation equilibrated the walker population, and sampled the 1-, 2-, and 3-body active space RDMs (12,000 iterations), which require finer sampling than the $\overline{G^1}$ sampling. This more expensive 4-body sampling is performed in the second phase of the calculation (200 iterations), where the stochastic promotion is used, with parameters $\chimin=10^{-7}$, $\chimax=10^{-4}$, $p(\chimin)=0.05$ and $p(\chimax)=0.95$. The sampling of the higher-body excitations was performed with a granularity of 5, meaning that one 3- or 4-body excitation was sampled for every five walkers, each iteration \cite{doi:10.1063/1.5140086}. Finally, deterministic core spaces of 10,000 determinants were used for the exact walker propagation to reduce random errors \cite{PhysRevLett.109.230201,doi:10.1063/1.4920975}.

As can be seen in the figure, the convergence of the FCIQMC-NEVPT2 energies is rapid, with convergence from only 500,000 walkers required for $\mathcal{O}[10^{-4}]E_{\rm h}$ accuracy in the energy. The 2 million walker calculation took only 4.3 hours on 32 compute cores (one node) for phase 1 of the simulation, with the $\overline{G^1}$ accumulation in phase 2 with the stochastic promotion scheme taking 5.1 hours for each seed. This time is reduced by approximately 80\% due to the stochastic promotion as compared to the previous fully deterministic promotion scheme. Similar convergence characteristics were found for the \quintet state. This relatively small computational effort indicates the potential for scaling up the size of these calculations to active spaces of larger systems with many correlated centres.

\begin{figure}[t]
\centering
   \includegraphics[width=\columnwidth]{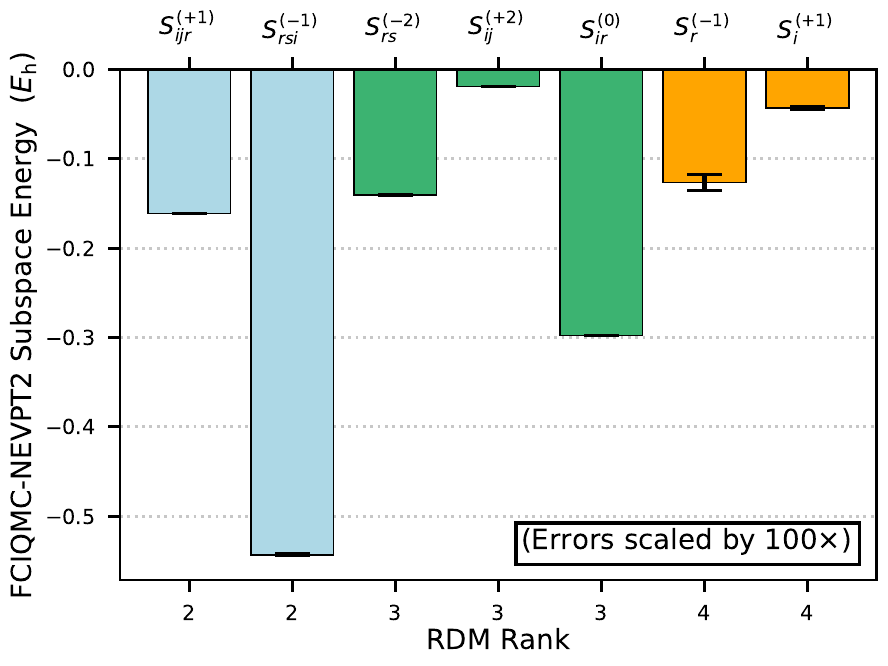}
	\caption{Breakdown of RDM-dependent sc-NEVPT2 energy contributions by perturber subspace for iron porphyrin ${}^1A_{1g}$ state. 
	Not shown is the dominant contribution which comes from the MP2-like perturber type which captures two electron scattering processes in the external space, and which does not directly depend on the CAS density matrices. This had a value of $-2.399544(3)\; E_\mathrm{h}$.
	As can be seen from the (scaled) error bars, the stochastic error is only at most $\mathcal{O}[10^{-5}]$ in each subspace, with the error in the $S_r^{(-1)}$ subspace dominating.}
   \label{fig:1A1g_barchart} 
\end{figure}

The best estimate of the FCIQMC CASSCF-NEVPT2 (24,24) energy of the \singlet state is found to be -2248.50696(9)$E_{\rm h}$, with the stochastic NEVPT2 contribution to this energy being -3.73055(9)$E_{\rm h}$. The number in parentheses indicates the estimated standard error in the previous digit from five independent calculations. \add{This random error is wholly derived from the stochastic NEVPT2, with the random error from the underlying FCIQMC-CASSCF reference negligible in comparison}. In comparison, the quintet state NEVPT2 correlation contribution is smaller, given by -3.6724(3)$E_{\rm h}$, indicating a preferential stabilisation of the singlet state over the quintet from the additional NEVPT2 treatment of correlation in the external space. It is expected that an improved correlation treatment will improve the lower spin states, and demonstrates the importance of an accurate correlated treatment in the study of spin gaps. Nevertheless, given the incompleteness of the basis set and neglect of even scalar relativistic effects, we do not consider these results to be a quantitative result, which is still forthcoming with this methodology.  

%\begin{subfigure}[b]{\columnwidth}
%   \includegraphics[width=\columnwidth]{figures/5A1g_nevpt2.pdf}
%   \caption{\quintet state}
%   \label{fig:5A1g_nevpt2}
%\end{subfigure}

%\end{figure}

As final insight into the approach, in Fig.~\ref{fig:1A1g_barchart} we present a breakdown for the different contributions to the final FCIQMC-NEVPT2 energy for the \singlet state, also denoting the effective maximum RDM rank which contributes. The $S_i^{(+1)}$, and particularly the $S_r^{(-1)}$ subspaces are dominant contributors to the overall random error in the FCIQMC-NEVPT2 results. This is to be expected, since they are the ones which rely on the $\overline{G^1}$ intermediate, and are sampled via the additional stochastic step proposed. Furthermore, their contributions are only sampled in the second, shorter phase of the simulation, since their cost still represents the largest of the calculation. However despite being sampled for a shorter number of iterations, their relative magnitude is not negligible, but nevertheless can still be efficiently sampled within the approach.

\section{Conclusions}
\label{sect:Conclusion}

We have detailed an improved algorithm for the calculation of stochastic FCIQMC-NEVPT2 energies. This algorithm introduces a nested stochastic step within the promotion of low-body excitations to high-rank intermediates required for the computation of the most costly $\mathcal{E}_{i}^{(+1)}$ and $\mathcal{E}_{r}^{(-1)}$ subspace energies in NEVPT2. Benchmarking this nested stochastic scheme on the \CrD system, we find a wide range of parameters for this stochastic sampling which avoids the introduction of any further error, but can reduce simulation time by up to 80\%. We outline an optimised simulation procedure, which attempts to minimise errors in the cheaper subspaces before the more expensive sampling of the $\mathcal{E}_{i}^{(+1)}$ and $\mathcal{E}_{r}^{(-1)}$ intermediates.

This enhanced sampling scheme is then applied to a (24,24) active space of iron porphyrin, combined with full stochastic orbital optimisation. Again, we find significant speedups through this scheme, and reliable convergence with respect to walker number. The \singlet spin state is found to be preferentially stabilised by the stochastic NEVPT2 treatment, in broad agreement with previous results, although quantitative accuracy will require a more careful treatment of remaining basis set incompleteness in this system. Overall simulation times for this active space size are found to be achievable within large workstation or small compute cluster resources, demonstrating the feasibility of FCIQMC-NEVPT2 with this adaptation for routine calculations in this size of active space. Future work will apply the approach to the consideration of larger active spaces, where FCIQMC has shown to still be effective with many millions of walkers used in a distribution fashion, as well as the computation of excited state FCIQMC-NEVPT2 in the same symmetry sector \cite{Blunt/2015,doi:10.1063/1.4766327,PhysRevLett.115.050603}.

\begin{acknowledgements} %%%
 The authors dedicate this paper in honour of Professor J{\" u}rgen Gauss on the occasion of his 60th birthday. G.H.B. gratefully acknowledges support from the Royal Society via a University Research Fellowship, as well as funding from the European Union's Horizon 2020 research and innovation programme under grant agreement No. 759063. We are also grateful to the UK Materials and Molecular Modelling Hub for computational resources, which is partially funded by EPSRC (EP/P020194/1).
\end{acknowledgements}

%\end{multicols}
%\bibliography{bibliography}{}

\begin{thebibliography}{45}%
\makeatletter
\providecommand \@ifxundefined [1]{%
 \@ifx{#1\undefined}
}%
\providecommand \@ifnum [1]{%
 \ifnum #1\expandafter \@firstoftwo
 \else \expandafter \@secondoftwo
 \fi
}%
\providecommand \@ifx [1]{%
 \ifx #1\expandafter \@firstoftwo
 \else \expandafter \@secondoftwo
 \fi
}%
\providecommand \natexlab [1]{#1}%
\providecommand \enquote  [1]{``#1''}%
\providecommand \bibnamefont  [1]{#1}%
\providecommand \bibfnamefont [1]{#1}%
\providecommand \citenamefont [1]{#1}%
\providecommand \href@noop [0]{\@secondoftwo}%
\providecommand \href [0]{\begingroup \@sanitize@url \@href}%
\providecommand \@href[1]{\@@startlink{#1}\@@href}%
\providecommand \@@href[1]{\endgroup#1\@@endlink}%
\providecommand \@sanitize@url [0]{\catcode `\\12\catcode `\$12\catcode
  `\&12\catcode `\#12\catcode `\^12\catcode `\_12\catcode `\%12\relax}%
\providecommand \@@startlink[1]{}%
\providecommand \@@endlink[0]{}%
\providecommand \url  [0]{\begingroup\@sanitize@url \@url }%
\providecommand \@url [1]{\endgroup\@href {#1}{\urlprefix }}%
\providecommand \urlprefix  [0]{URL }%
\providecommand \Eprint [0]{\href }%
\providecommand \doibase [0]{http://dx.doi.org/}%
\providecommand \selectlanguage [0]{\@gobble}%
\providecommand \bibinfo  [0]{\@secondoftwo}%
\providecommand \bibfield  [0]{\@secondoftwo}%
\providecommand \translation [1]{[#1]}%
\providecommand \BibitemOpen [0]{}%
\providecommand \bibitemStop [0]{}%
\providecommand \bibitemNoStop [0]{.\EOS\space}%
\providecommand \EOS [0]{\spacefactor3000\relax}%
\providecommand \BibitemShut  [1]{\csname bibitem#1\endcsname}%
\let\auto@bib@innerbib\@empty
%</preamble>
\bibitem [{\citenamefont {Booth}\ \emph
  {et~al.}(2012{\natexlab{a}})\citenamefont {Booth}, \citenamefont
  {Gr{\"u}neis}, \citenamefont {Kresse},\ and\ \citenamefont
  {Alavi}}]{booth/nature}%
  \BibitemOpen
  \bibfield  {author} {\bibinfo {author} {\bibfnamefont {G.~H.}\ \bibnamefont
  {Booth}}, \bibinfo {author} {\bibfnamefont {A.}~\bibnamefont {Gr{\"u}neis}},
  \bibinfo {author} {\bibfnamefont {G.}~\bibnamefont {Kresse}}, \ and\ \bibinfo
  {author} {\bibfnamefont {A.}~\bibnamefont {Alavi}},\ }\href
  {https://doi.org/10.1038/nature11770} {\bibfield  {journal} {\bibinfo
  {journal} {Nature}\ }\textbf {\bibinfo {volume} {493}},\ \bibinfo {pages}
  {365} (\bibinfo {year} {2012}{\natexlab{a}})}\BibitemShut {NoStop}%
\bibitem [{\citenamefont {Booth}, \citenamefont {Thom},\ and\ \citenamefont
  {Alavi}(2009)}]{doi:10.1063/1.3193710}%
  \BibitemOpen
  \bibfield  {author} {\bibinfo {author} {\bibfnamefont {G.~H.}\ \bibnamefont
  {Booth}}, \bibinfo {author} {\bibfnamefont {A.~J.~W.}\ \bibnamefont {Thom}},
  \ and\ \bibinfo {author} {\bibfnamefont {A.}~\bibnamefont {Alavi}},\ }\href
  {\doibase 10.1063/1.3193710} {\bibfield  {journal} {\bibinfo  {journal} {J.
  Chem. Phys.}\ }\textbf {\bibinfo {volume} {131}},\ \bibinfo {pages} {054106}
  (\bibinfo {year} {2009})}\BibitemShut {NoStop}%
\bibitem [{\citenamefont {Spencer}, \citenamefont {Blunt},\ and\ \citenamefont
  {Foulkes}(2012)}]{Spencer_2012}%
  \BibitemOpen
  \bibfield  {author} {\bibinfo {author} {\bibfnamefont {J.~S.}\ \bibnamefont
  {Spencer}}, \bibinfo {author} {\bibfnamefont {N.~S.}\ \bibnamefont {Blunt}},
  \ and\ \bibinfo {author} {\bibfnamefont {W.~M.}\ \bibnamefont {Foulkes}},\
  }\href {\doibase 10.1063/1.3681396} {\bibfield  {journal} {\bibinfo
  {journal} {J. Chem. Phys.}\ }\textbf {\bibinfo {volume} {136}},\ \bibinfo
  {pages} {054110} (\bibinfo {year} {2012})}\BibitemShut {NoStop}%
\bibitem [{\citenamefont {Cleland}, \citenamefont {Booth},\ and\ \citenamefont
  {Alavi}(2010)}]{doi:10.1063/1.3302277}%
  \BibitemOpen
  \bibfield  {author} {\bibinfo {author} {\bibfnamefont {D.}~\bibnamefont
  {Cleland}}, \bibinfo {author} {\bibfnamefont {G.~H.}\ \bibnamefont {Booth}},
  \ and\ \bibinfo {author} {\bibfnamefont {A.}~\bibnamefont {Alavi}},\ }\href
  {\doibase 10.1063/1.3302277} {\bibfield  {journal} {\bibinfo  {journal} {J.
  Chem. Phys.}\ }\textbf {\bibinfo {volume} {132}},\ \bibinfo {pages} {041103}
  (\bibinfo {year} {2010})}\BibitemShut {NoStop}%
\bibitem [{\citenamefont {Blunt}\ \emph
  {et~al.}(2015{\natexlab{a}})\citenamefont {Blunt}, \citenamefont {Smart},
  \citenamefont {Booth},\ and\ \citenamefont {Alavi}}]{Blunt/2015}%
  \BibitemOpen
  \bibfield  {author} {\bibinfo {author} {\bibfnamefont {N.~S.}\ \bibnamefont
  {Blunt}}, \bibinfo {author} {\bibfnamefont {S.~D.}\ \bibnamefont {Smart}},
  \bibinfo {author} {\bibfnamefont {G.~H.}\ \bibnamefont {Booth}}, \ and\
  \bibinfo {author} {\bibfnamefont {A.}~\bibnamefont {Alavi}},\ }\href
  {\doibase 10.1063/1.4932595} {\bibfield  {journal} {\bibinfo  {journal} {J.
  Chem. Phys.}\ }\textbf {\bibinfo {volume} {143}},\ \bibinfo {pages} {134117}
  (\bibinfo {year} {2015}{\natexlab{a}})}\BibitemShut {NoStop}%
\bibitem [{\citenamefont {Thomas}, \citenamefont {Booth},\ and\ \citenamefont
  {Alavi}(2015)}]{PhysRevLett.114.033001}%
  \BibitemOpen
  \bibfield  {author} {\bibinfo {author} {\bibfnamefont {R.~E.}\ \bibnamefont
  {Thomas}}, \bibinfo {author} {\bibfnamefont {G.~H.}\ \bibnamefont {Booth}}, \
  and\ \bibinfo {author} {\bibfnamefont {A.}~\bibnamefont {Alavi}},\ }\href
  {\doibase 10.1103/PhysRevLett.114.033001} {\bibfield  {journal} {\bibinfo
  {journal} {Phys. Rev. Lett.}\ }\textbf {\bibinfo {volume} {114}},\ \bibinfo
  {pages} {033001} (\bibinfo {year} {2015})}\BibitemShut {NoStop}%
\bibitem [{\citenamefont {Petruzielo}\ \emph {et~al.}(2012)\citenamefont
  {Petruzielo}, \citenamefont {Holmes}, \citenamefont {Changlani},
  \citenamefont {Nightingale},\ and\ \citenamefont
  {Umrigar}}]{PhysRevLett.109.230201}%
  \BibitemOpen
  \bibfield  {author} {\bibinfo {author} {\bibfnamefont {F.~R.}\ \bibnamefont
  {Petruzielo}}, \bibinfo {author} {\bibfnamefont {A.~A.}\ \bibnamefont
  {Holmes}}, \bibinfo {author} {\bibfnamefont {H.~J.}\ \bibnamefont
  {Changlani}}, \bibinfo {author} {\bibfnamefont {M.~P.}\ \bibnamefont
  {Nightingale}}, \ and\ \bibinfo {author} {\bibfnamefont {C.~J.}\ \bibnamefont
  {Umrigar}},\ }\href {\doibase 10.1103/PhysRevLett.109.230201} {\bibfield
  {journal} {\bibinfo  {journal} {Phys. Rev. Lett.}\ }\textbf {\bibinfo
  {volume} {109}},\ \bibinfo {pages} {230201} (\bibinfo {year}
  {2012})}\BibitemShut {NoStop}%
\bibitem [{\citenamefont {Williams}\ \emph {et~al.}(2020)\citenamefont
  {Williams}, \citenamefont {Yao}, \citenamefont {Li}, \citenamefont {Chen},
  \citenamefont {Shi}, \citenamefont {Motta}, \citenamefont {Niu},
  \citenamefont {Ray}, \citenamefont {Guo}, \citenamefont {Anderson},
  \citenamefont {Li}, \citenamefont {Tran}, \citenamefont {Yeh}, \citenamefont
  {Mussard}, \citenamefont {Sharma}, \citenamefont {Bruneval}, \citenamefont
  {van Schilfgaarde}, \citenamefont {Booth}, \citenamefont {Chan},
  \citenamefont {Zhang}, \citenamefont {Gull}, \citenamefont {Zgid},
  \citenamefont {Millis}, \citenamefont {Umrigar},\ and\ \citenamefont
  {Wagner}}]{PhysRevX.10.011041}%
  \BibitemOpen
  \bibfield  {author} {\bibinfo {author} {\bibfnamefont {K.~T.}\ \bibnamefont
  {Williams}}, \bibinfo {author} {\bibfnamefont {Y.}~\bibnamefont {Yao}},
  \bibinfo {author} {\bibfnamefont {J.}~\bibnamefont {Li}}, \bibinfo {author}
  {\bibfnamefont {L.}~\bibnamefont {Chen}}, \bibinfo {author} {\bibfnamefont
  {H.}~\bibnamefont {Shi}}, \bibinfo {author} {\bibfnamefont {M.}~\bibnamefont
  {Motta}}, \bibinfo {author} {\bibfnamefont {C.}~\bibnamefont {Niu}}, \bibinfo
  {author} {\bibfnamefont {U.}~\bibnamefont {Ray}}, \bibinfo {author}
  {\bibfnamefont {S.}~\bibnamefont {Guo}}, \bibinfo {author} {\bibfnamefont
  {R.~J.}\ \bibnamefont {Anderson}}, \bibinfo {author} {\bibfnamefont
  {J.}~\bibnamefont {Li}}, \bibinfo {author} {\bibfnamefont {L.~N.}\
  \bibnamefont {Tran}}, \bibinfo {author} {\bibfnamefont {C.-N.}\ \bibnamefont
  {Yeh}}, \bibinfo {author} {\bibfnamefont {B.}~\bibnamefont {Mussard}},
  \bibinfo {author} {\bibfnamefont {S.}~\bibnamefont {Sharma}}, \bibinfo
  {author} {\bibfnamefont {F.}~\bibnamefont {Bruneval}}, \bibinfo {author}
  {\bibfnamefont {M.}~\bibnamefont {van Schilfgaarde}}, \bibinfo {author}
  {\bibfnamefont {G.~H.}\ \bibnamefont {Booth}}, \bibinfo {author}
  {\bibfnamefont {G.~K.-L.}\ \bibnamefont {Chan}}, \bibinfo {author}
  {\bibfnamefont {S.}~\bibnamefont {Zhang}}, \bibinfo {author} {\bibfnamefont
  {E.}~\bibnamefont {Gull}}, \bibinfo {author} {\bibfnamefont {D.}~\bibnamefont
  {Zgid}}, \bibinfo {author} {\bibfnamefont {A.}~\bibnamefont {Millis}},
  \bibinfo {author} {\bibfnamefont {C.~J.}\ \bibnamefont {Umrigar}}, \ and\
  \bibinfo {author} {\bibfnamefont {L.~K.}\ \bibnamefont {Wagner}} (\bibinfo
  {collaboration} {Simons Collaboration on the Many-Electron Problem}),\ }\href
  {\doibase 10.1103/PhysRevX.10.011041} {\bibfield  {journal} {\bibinfo
  {journal} {Phys. Rev. X}\ }\textbf {\bibinfo {volume} {10}},\ \bibinfo
  {pages} {011041} (\bibinfo {year} {2020})}\BibitemShut {NoStop}%
\bibitem [{\citenamefont {Bytautas}\ and\ \citenamefont
  {Ruedenberg}(2009)}]{BYTAUTAS200964}%
  \BibitemOpen
  \bibfield  {author} {\bibinfo {author} {\bibfnamefont {L.}~\bibnamefont
  {Bytautas}}\ and\ \bibinfo {author} {\bibfnamefont {K.}~\bibnamefont
  {Ruedenberg}},\ }\href {\doibase
  https://doi.org/10.1016/j.chemphys.2008.11.021} {\bibfield  {journal}
  {\bibinfo  {journal} {Chem. Phys.}\ }\textbf {\bibinfo {volume} {356}},\
  \bibinfo {pages} {64 } (\bibinfo {year} {2009})}\BibitemShut {NoStop}%
\bibitem [{\citenamefont {Booth}\ \emph {et~al.}(2011)\citenamefont {Booth},
  \citenamefont {Cleland}, \citenamefont {Thom},\ and\ \citenamefont
  {Alavi}}]{doi:10.1063/1.3624383}%
  \BibitemOpen
  \bibfield  {author} {\bibinfo {author} {\bibfnamefont {G.~H.}\ \bibnamefont
  {Booth}}, \bibinfo {author} {\bibfnamefont {D.}~\bibnamefont {Cleland}},
  \bibinfo {author} {\bibfnamefont {A.~J.~W.}\ \bibnamefont {Thom}}, \ and\
  \bibinfo {author} {\bibfnamefont {A.}~\bibnamefont {Alavi}},\ }\href
  {\doibase 10.1063/1.3624383} {\bibfield  {journal} {\bibinfo  {journal} {J.
  Chem. Phys.}\ }\textbf {\bibinfo {volume} {135}},\ \bibinfo {pages} {084104}
  (\bibinfo {year} {2011})}\BibitemShut {NoStop}%
\bibitem [{\citenamefont {Anderson}, \citenamefont {Shiozaki},\ and\
  \citenamefont {Booth}(2020)}]{doi:10.1063/1.5140086}%
  \BibitemOpen
  \bibfield  {author} {\bibinfo {author} {\bibfnamefont {R.~J.}\ \bibnamefont
  {Anderson}}, \bibinfo {author} {\bibfnamefont {T.}~\bibnamefont {Shiozaki}},
  \ and\ \bibinfo {author} {\bibfnamefont {G.~H.}\ \bibnamefont {Booth}},\
  }\href {\doibase 10.1063/1.5140086} {\bibfield  {journal} {\bibinfo
  {journal} {J. Chem. Phys.}\ }\textbf {\bibinfo {volume} {152}},\ \bibinfo
  {pages} {054101} (\bibinfo {year} {2020})}\BibitemShut {NoStop}%
\bibitem [{\citenamefont {Overy}\ \emph {et~al.}(2014)\citenamefont {Overy},
  \citenamefont {Booth}, \citenamefont {Blunt}, \citenamefont {Shepherd},
  \citenamefont {Cleland},\ and\ \citenamefont
  {Alavi}}]{doi:10.1063/1.4904313}%
  \BibitemOpen
  \bibfield  {author} {\bibinfo {author} {\bibfnamefont {C.}~\bibnamefont
  {Overy}}, \bibinfo {author} {\bibfnamefont {G.}~\bibnamefont {Booth}},
  \bibinfo {author} {\bibfnamefont {N.}~\bibnamefont {Blunt}}, \bibinfo
  {author} {\bibfnamefont {J.}~\bibnamefont {Shepherd}}, \bibinfo {author}
  {\bibfnamefont {D.}~\bibnamefont {Cleland}}, \ and\ \bibinfo {author}
  {\bibfnamefont {A.}~\bibnamefont {Alavi}},\ }\href@noop {} {\bibfield
  {journal} {\bibinfo  {journal} {J. Chem. Phys.}\ }\textbf {\bibinfo {volume}
  {141}},\ \bibinfo {pages} {244117} (\bibinfo {year} {2014})}\BibitemShut
  {NoStop}%
\bibitem [{\citenamefont {Jeanmairet}, \citenamefont {Sharma},\ and\
  \citenamefont {Alavi}(2017)}]{doi:10.1063/1.4974177}%
  \BibitemOpen
  \bibfield  {author} {\bibinfo {author} {\bibfnamefont {G.}~\bibnamefont
  {Jeanmairet}}, \bibinfo {author} {\bibfnamefont {S.}~\bibnamefont {Sharma}},
  \ and\ \bibinfo {author} {\bibfnamefont {A.}~\bibnamefont {Alavi}},\ }\href
  {\doibase 10.1063/1.4974177} {\bibfield  {journal} {\bibinfo  {journal} {J.
  Chem. Phys.}\ }\textbf {\bibinfo {volume} {146}},\ \bibinfo {pages} {044107}
  (\bibinfo {year} {2017})}\BibitemShut {NoStop}%
\bibitem [{\citenamefont {Andersson}, \citenamefont {Malmqvist},\ and\
  \citenamefont {Roos}(1992)}]{doi:10.1063/1.462209}%
  \BibitemOpen
  \bibfield  {author} {\bibinfo {author} {\bibfnamefont {K.}~\bibnamefont
  {Andersson}}, \bibinfo {author} {\bibfnamefont {P.~A.}\ \bibnamefont
  {Malmqvist}}, \ and\ \bibinfo {author} {\bibfnamefont {B.~O.}\ \bibnamefont
  {Roos}},\ }\href {\doibase 10.1063/1.462209} {\bibfield  {journal} {\bibinfo
  {journal} {J. Chem. Phys.}\ }\textbf {\bibinfo {volume} {96}},\ \bibinfo
  {pages} {1218} (\bibinfo {year} {1992})}\BibitemShut {NoStop}%
\bibitem [{\citenamefont {Zobel}, \citenamefont {Nogueira},\ and\ \citenamefont
  {González}(2017)}]{C6SC03759C}%
  \BibitemOpen
  \bibfield  {author} {\bibinfo {author} {\bibfnamefont {J.~P.}\ \bibnamefont
  {Zobel}}, \bibinfo {author} {\bibfnamefont {J.~J.}\ \bibnamefont {Nogueira}},
  \ and\ \bibinfo {author} {\bibfnamefont {L.}~\bibnamefont {González}},\
  }\href {\doibase 10.1039/C6SC03759C} {\bibfield  {journal} {\bibinfo
  {journal} {Chem. Sci.}\ }\textbf {\bibinfo {volume} {8}},\ \bibinfo {pages}
  {1482} (\bibinfo {year} {2017})}\BibitemShut {NoStop}%
\bibitem [{\citenamefont {Thomas}\ \emph {et~al.}(2015)\citenamefont {Thomas},
  \citenamefont {Sun}, \citenamefont {Alavi},\ and\ \citenamefont
  {Booth}}]{doi:10.1021/acs.jctc.5b00917}%
  \BibitemOpen
  \bibfield  {author} {\bibinfo {author} {\bibfnamefont {R.~E.}\ \bibnamefont
  {Thomas}}, \bibinfo {author} {\bibfnamefont {Q.}~\bibnamefont {Sun}},
  \bibinfo {author} {\bibfnamefont {A.}~\bibnamefont {Alavi}}, \ and\ \bibinfo
  {author} {\bibfnamefont {G.~H.}\ \bibnamefont {Booth}},\ }\href {\doibase
  10.1021/acs.jctc.5b00917} {\bibfield  {journal} {\bibinfo  {journal} {J.
  Chem. Theory Comput.}\ }\textbf {\bibinfo {volume} {11}},\ \bibinfo {pages}
  {5316} (\bibinfo {year} {2015})}\BibitemShut {NoStop}%
\bibitem [{\citenamefont {Booth}, \citenamefont {Smart},\ and\ \citenamefont
  {Alavi}(2014)}]{booth_fciqmc_review}%
  \BibitemOpen
  \bibfield  {author} {\bibinfo {author} {\bibfnamefont {G.~H.}\ \bibnamefont
  {Booth}}, \bibinfo {author} {\bibfnamefont {S.}~\bibnamefont {Smart}}, \ and\
  \bibinfo {author} {\bibfnamefont {A.}~\bibnamefont {Alavi}},\ }\href
  {\doibase 10.1080/00268976.2013.877165} {\bibfield  {journal} {\bibinfo
  {journal} {Mol. Phys.}\ }\textbf {\bibinfo {volume} {112}},\ \bibinfo {pages}
  {1855} (\bibinfo {year} {2014})}\BibitemShut {NoStop}%
\bibitem [{\citenamefont {Holmes}, \citenamefont {Changlani},\ and\
  \citenamefont {Umrigar}(2016)}]{doi:10.1021/acs.jctc.5b01170}%
  \BibitemOpen
  \bibfield  {author} {\bibinfo {author} {\bibfnamefont {A.~A.}\ \bibnamefont
  {Holmes}}, \bibinfo {author} {\bibfnamefont {H.~J.}\ \bibnamefont
  {Changlani}}, \ and\ \bibinfo {author} {\bibfnamefont {C.~J.}\ \bibnamefont
  {Umrigar}},\ }\href {\doibase 10.1021/acs.jctc.5b01170} {\bibfield  {journal}
  {\bibinfo  {journal} {J. Chem. Theor. Comput.}\ }\textbf {\bibinfo {volume}
  {12}},\ \bibinfo {pages} {1561} (\bibinfo {year} {2016})}\BibitemShut
  {NoStop}%
\bibitem [{\citenamefont {Booth}\ \emph
  {et~al.}(2012{\natexlab{b}})\citenamefont {Booth}, \citenamefont {Cleland},
  \citenamefont {Alavi},\ and\ \citenamefont {Tew}}]{doi:10.1063/1.4762445}%
  \BibitemOpen
  \bibfield  {author} {\bibinfo {author} {\bibfnamefont {G.~H.}\ \bibnamefont
  {Booth}}, \bibinfo {author} {\bibfnamefont {D.}~\bibnamefont {Cleland}},
  \bibinfo {author} {\bibfnamefont {A.}~\bibnamefont {Alavi}}, \ and\ \bibinfo
  {author} {\bibfnamefont {D.~P.}\ \bibnamefont {Tew}},\ }\href {\doibase
  10.1063/1.4762445} {\bibfield  {journal} {\bibinfo  {journal} {J. Chem.
  Phys.}\ }\textbf {\bibinfo {volume} {137}},\ \bibinfo {pages} {164112}
  (\bibinfo {year} {2012}{\natexlab{b}})}\BibitemShut {NoStop}%
\bibitem [{\citenamefont {Blunt}, \citenamefont {Booth},\ and\ \citenamefont
  {Alavi}(2017)}]{doi:10.1063/1.4986963}%
  \BibitemOpen
  \bibfield  {author} {\bibinfo {author} {\bibfnamefont {N.~S.}\ \bibnamefont
  {Blunt}}, \bibinfo {author} {\bibfnamefont {G.~H.}\ \bibnamefont {Booth}}, \
  and\ \bibinfo {author} {\bibfnamefont {A.}~\bibnamefont {Alavi}},\ }\href
  {\doibase 10.1063/1.4986963} {\bibfield  {journal} {\bibinfo  {journal} {J.
  Chem. Phys.}\ }\textbf {\bibinfo {volume} {146}},\ \bibinfo {pages} {244105}
  (\bibinfo {year} {2017})}\BibitemShut {NoStop}%
\bibitem [{\citenamefont {Finley}\ \emph {et~al.}(1998)\citenamefont {Finley},
  \citenamefont {Malmqvist}, \citenamefont {Roos},\ and\ \citenamefont
  {Serrano-Andrés}}]{FINLEY1998299}%
  \BibitemOpen
  \bibfield  {author} {\bibinfo {author} {\bibfnamefont {J.}~\bibnamefont
  {Finley}}, \bibinfo {author} {\bibfnamefont {P.-A.}\ \bibnamefont
  {Malmqvist}}, \bibinfo {author} {\bibfnamefont {B.~O.}\ \bibnamefont {Roos}},
  \ and\ \bibinfo {author} {\bibfnamefont {L.}~\bibnamefont
  {Serrano-Andrés}},\ }\href {\doibase
  https://doi.org/10.1016/S0009-2614(98)00252-8} {\bibfield  {journal}
  {\bibinfo  {journal} {Chem. Phys. Lett.}\ }\textbf {\bibinfo {volume}
  {288}},\ \bibinfo {pages} {299 } (\bibinfo {year} {1998})}\BibitemShut
  {NoStop}%
\bibitem [{\citenamefont {Angeli}, \citenamefont {Cimiraglia},\ and\
  \citenamefont {Malrieu}(2001)}]{ANGELI2001297}%
  \BibitemOpen
  \bibfield  {author} {\bibinfo {author} {\bibfnamefont {C.}~\bibnamefont
  {Angeli}}, \bibinfo {author} {\bibfnamefont {R.}~\bibnamefont {Cimiraglia}},
  \ and\ \bibinfo {author} {\bibfnamefont {J.-P.}\ \bibnamefont {Malrieu}},\
  }\href@noop {} {\bibfield  {journal} {\bibinfo  {journal} {Chem. Phys.
  Lett.}\ }\textbf {\bibinfo {volume} {350}},\ \bibinfo {pages} {297 }
  (\bibinfo {year} {2001})}\BibitemShut {NoStop}%
\bibitem [{\citenamefont {Andersson}\ \emph {et~al.}(1990)\citenamefont
  {Andersson}, \citenamefont {Malmqvist}, \citenamefont {Roos}, \citenamefont
  {Sadlej},\ and\ \citenamefont {Wolinski}}]{doi:10.1021/j100377a012}%
  \BibitemOpen
  \bibfield  {author} {\bibinfo {author} {\bibfnamefont {K.}~\bibnamefont
  {Andersson}}, \bibinfo {author} {\bibfnamefont {P.~A.}\ \bibnamefont
  {Malmqvist}}, \bibinfo {author} {\bibfnamefont {B.~O.}\ \bibnamefont {Roos}},
  \bibinfo {author} {\bibfnamefont {A.~J.}\ \bibnamefont {Sadlej}}, \ and\
  \bibinfo {author} {\bibfnamefont {K.}~\bibnamefont {Wolinski}},\ }\href
  {\doibase 10.1021/j100377a012} {\bibfield  {journal} {\bibinfo  {journal} {J.
  Phys. Chem.}\ }\textbf {\bibinfo {volume} {94}},\ \bibinfo {pages} {5483}
  (\bibinfo {year} {1990})}\BibitemShut {NoStop}%
\bibitem [{\citenamefont {Blunt}, \citenamefont {Alavi},\ and\ \citenamefont
  {Booth}(2018)}]{PhysRevB.98.085118}%
  \BibitemOpen
  \bibfield  {author} {\bibinfo {author} {\bibfnamefont {N.~S.}\ \bibnamefont
  {Blunt}}, \bibinfo {author} {\bibfnamefont {A.}~\bibnamefont {Alavi}}, \ and\
  \bibinfo {author} {\bibfnamefont {G.~H.}\ \bibnamefont {Booth}},\ }\href
  {\doibase 10.1103/PhysRevB.98.085118} {\bibfield  {journal} {\bibinfo
  {journal} {Phys. Rev. B}\ }\textbf {\bibinfo {volume} {98}},\ \bibinfo
  {pages} {085118} (\bibinfo {year} {2018})}\BibitemShut {NoStop}%
\bibitem [{\citenamefont {Angeli}\ \emph {et~al.}(2001)\citenamefont {Angeli},
  \citenamefont {Cimiraglia}, \citenamefont {Evangelisti}, \citenamefont
  {Leininger},\ and\ \citenamefont {Malrieu}}]{doi:10.1063/1.1361246}%
  \BibitemOpen
  \bibfield  {author} {\bibinfo {author} {\bibfnamefont {C.}~\bibnamefont
  {Angeli}}, \bibinfo {author} {\bibfnamefont {R.}~\bibnamefont {Cimiraglia}},
  \bibinfo {author} {\bibfnamefont {S.}~\bibnamefont {Evangelisti}}, \bibinfo
  {author} {\bibfnamefont {T.}~\bibnamefont {Leininger}}, \ and\ \bibinfo
  {author} {\bibfnamefont {J.-P.}\ \bibnamefont {Malrieu}},\ }\href {\doibase
  10.1063/1.1361246} {\bibfield  {journal} {\bibinfo  {journal} {J. Chem.
  Phys.}\ }\textbf {\bibinfo {volume} {114}},\ \bibinfo {pages} {10252}
  (\bibinfo {year} {2001})}\BibitemShut {NoStop}%
\bibitem [{\citenamefont {Angeli}, \citenamefont {Cimiraglia},\ and\
  \citenamefont {Malrieu}(2002)}]{doi:10.1063/1.1515317}%
  \BibitemOpen
  \bibfield  {author} {\bibinfo {author} {\bibfnamefont {C.}~\bibnamefont
  {Angeli}}, \bibinfo {author} {\bibfnamefont {R.}~\bibnamefont {Cimiraglia}},
  \ and\ \bibinfo {author} {\bibfnamefont {J.-P.}\ \bibnamefont {Malrieu}},\
  }\href {\doibase 10.1063/1.1515317} {\bibfield  {journal} {\bibinfo
  {journal} {J. Chem. Phys.}\ }\textbf {\bibinfo {volume} {117}},\ \bibinfo
  {pages} {9138} (\bibinfo {year} {2002})}\BibitemShut {NoStop}%
\bibitem [{Note1()}]{Note1}%
  \BibitemOpen
  \bibinfo {note} {In Fig.~\ref {fig:chi_example}, triple and quadruple
  excitations are sampled frequently relative to singles and doubles. These
  higher two excitation levels are sampled purely for the purposes of
  higher-body RDM sampling and do not affect the dynamics. Therefore, their
  relative ratio of sampling compared to single and double excitations is not a
  physical quantity, but rather a user-defined choice. This is controlled by a
  `granularity' parameter (detailed in previous work \cite
  {doi:10.1063/1.5140086}), which in this example was set to one, ensuring
  significant sampling of the higher-body excitations. In production
  simulations, we often choose to sample the higher-rank excitations less
  frequently. In this case the promotion of the more expensive single and
  double excitations increasingly dominate the cost, and the issue of
  mitigating their undesirable scaling becomes even more
  important.}\BibitemShut {Stop}%
\bibitem [{\citenamefont {Blunt}\ \emph
  {et~al.}(2015{\natexlab{b}})\citenamefont {Blunt}, \citenamefont {Smart},
  \citenamefont {Kersten}, \citenamefont {Spencer}, \citenamefont {Booth},\
  and\ \citenamefont {Alavi}}]{doi:10.1063/1.4920975}%
  \BibitemOpen
  \bibfield  {author} {\bibinfo {author} {\bibfnamefont {N.~S.}\ \bibnamefont
  {Blunt}}, \bibinfo {author} {\bibfnamefont {S.~D.}\ \bibnamefont {Smart}},
  \bibinfo {author} {\bibfnamefont {J.~A.~F.}\ \bibnamefont {Kersten}},
  \bibinfo {author} {\bibfnamefont {J.~S.}\ \bibnamefont {Spencer}}, \bibinfo
  {author} {\bibfnamefont {G.~H.}\ \bibnamefont {Booth}}, \ and\ \bibinfo
  {author} {\bibfnamefont {A.}~\bibnamefont {Alavi}},\ }\href {\doibase
  10.1063/1.4920975} {\bibfield  {journal} {\bibinfo  {journal} {J. Chem.
  Phys.}\ }\textbf {\bibinfo {volume} {142}},\ \bibinfo {pages} {184107}
  (\bibinfo {year} {2015}{\natexlab{b}})}\BibitemShut {NoStop}%
\bibitem [{\citenamefont {Sun}\ \emph {et~al.}(2017)\citenamefont {Sun},
  \citenamefont {Berkelbach}, \citenamefont {Blunt}, \citenamefont {Booth},
  \citenamefont {Guo}, \citenamefont {Li}, \citenamefont {Liu}, \citenamefont
  {McClain}, \citenamefont {Sayfutyarova}, \citenamefont {Sharma},
  \citenamefont {Wouters},\ and\ \citenamefont {Chan}}]{pyscf}%
  \BibitemOpen
  \bibfield  {author} {\bibinfo {author} {\bibfnamefont {Q.}~\bibnamefont
  {Sun}}, \bibinfo {author} {\bibfnamefont {T.~C.}\ \bibnamefont {Berkelbach}},
  \bibinfo {author} {\bibfnamefont {N.~S.}\ \bibnamefont {Blunt}}, \bibinfo
  {author} {\bibfnamefont {G.~H.}\ \bibnamefont {Booth}}, \bibinfo {author}
  {\bibfnamefont {S.}~\bibnamefont {Guo}}, \bibinfo {author} {\bibfnamefont
  {Z.}~\bibnamefont {Li}}, \bibinfo {author} {\bibfnamefont {J.}~\bibnamefont
  {Liu}}, \bibinfo {author} {\bibfnamefont {J.~D.}\ \bibnamefont {McClain}},
  \bibinfo {author} {\bibfnamefont {E.~R.}\ \bibnamefont {Sayfutyarova}},
  \bibinfo {author} {\bibfnamefont {S.}~\bibnamefont {Sharma}}, \bibinfo
  {author} {\bibfnamefont {S.}~\bibnamefont {Wouters}}, \ and\ \bibinfo
  {author} {\bibfnamefont {G.~K.-L.}\ \bibnamefont {Chan}},\ }\href {\doibase
  10.1002/wcms.1340} {\enquote {\bibinfo {title} {Pyscf: the python-based
  simulations of chemistry framework},}\ } (\bibinfo {year} {2017}),\ \Eprint
  {http://arxiv.org/abs/https://onlinelibrary.wiley.com/doi/pdf/10.1002/wcms.1340}
  {https://onlinelibrary.wiley.com/doi/pdf/10.1002/wcms.1340} \BibitemShut
  {NoStop}%
\bibitem [{\citenamefont {Li}\ \emph {et~al.}(2020)\citenamefont {Li},
  \citenamefont {Yao}, \citenamefont {Holmes}, \citenamefont {Otten},
  \citenamefont {Sun}, \citenamefont {Sharma},\ and\ \citenamefont
  {Umrigar}}]{PhysRevResearch.2.012015}%
  \BibitemOpen
  \bibfield  {author} {\bibinfo {author} {\bibfnamefont {J.}~\bibnamefont
  {Li}}, \bibinfo {author} {\bibfnamefont {Y.}~\bibnamefont {Yao}}, \bibinfo
  {author} {\bibfnamefont {A.~A.}\ \bibnamefont {Holmes}}, \bibinfo {author}
  {\bibfnamefont {M.}~\bibnamefont {Otten}}, \bibinfo {author} {\bibfnamefont
  {Q.}~\bibnamefont {Sun}}, \bibinfo {author} {\bibfnamefont {S.}~\bibnamefont
  {Sharma}}, \ and\ \bibinfo {author} {\bibfnamefont {C.~J.}\ \bibnamefont
  {Umrigar}},\ }\href {\doibase 10.1103/PhysRevResearch.2.012015} {\bibfield
  {journal} {\bibinfo  {journal} {Phys. Rev. Research}\ }\textbf {\bibinfo
  {volume} {2}},\ \bibinfo {pages} {012015} (\bibinfo {year}
  {2020})}\BibitemShut {NoStop}%
\bibitem [{\citenamefont {Purwanto}, \citenamefont {Zhang},\ and\ \citenamefont
  {Krakauer}(2015)}]{doi:10.1063/1.4906829}%
  \BibitemOpen
  \bibfield  {author} {\bibinfo {author} {\bibfnamefont {W.}~\bibnamefont
  {Purwanto}}, \bibinfo {author} {\bibfnamefont {S.}~\bibnamefont {Zhang}}, \
  and\ \bibinfo {author} {\bibfnamefont {H.}~\bibnamefont {Krakauer}},\ }\href
  {\doibase 10.1063/1.4906829} {\bibfield  {journal} {\bibinfo  {journal} {J.
  Chem. Phys.}\ }\textbf {\bibinfo {volume} {142}},\ \bibinfo {pages} {064302}
  (\bibinfo {year} {2015})}\BibitemShut {NoStop}%
\bibitem [{\citenamefont {Blunt}, \citenamefont {Alavi},\ and\ \citenamefont
  {Booth}(2015)}]{PhysRevLett.115.050603}%
  \BibitemOpen
  \bibfield  {author} {\bibinfo {author} {\bibfnamefont {N.~S.}\ \bibnamefont
  {Blunt}}, \bibinfo {author} {\bibfnamefont {A.}~\bibnamefont {Alavi}}, \ and\
  \bibinfo {author} {\bibfnamefont {G.~H.}\ \bibnamefont {Booth}},\ }\href
  {\doibase 10.1103/PhysRevLett.115.050603} {\bibfield  {journal} {\bibinfo
  {journal} {Phys. Rev. Lett.}\ }\textbf {\bibinfo {volume} {115}},\ \bibinfo
  {pages} {050603} (\bibinfo {year} {2015})}\BibitemShut {NoStop}%
\bibitem [{\citenamefont {Phung}, \citenamefont {Wouters},\ and\ \citenamefont
  {Pierloot}(2016)}]{doi:10.1021/acs.jctc.6b00714}%
  \BibitemOpen
  \bibfield  {author} {\bibinfo {author} {\bibfnamefont {Q.~M.}\ \bibnamefont
  {Phung}}, \bibinfo {author} {\bibfnamefont {S.}~\bibnamefont {Wouters}}, \
  and\ \bibinfo {author} {\bibfnamefont {K.}~\bibnamefont {Pierloot}},\ }\href
  {\doibase 10.1021/acs.jctc.6b00714} {\bibfield  {journal} {\bibinfo
  {journal} {J. Chem. Theory Comput.}\ }\textbf {\bibinfo {volume} {12}},\
  \bibinfo {pages} {4352} (\bibinfo {year} {2016})}\BibitemShut {NoStop}%
\bibitem [{\citenamefont {Li~Manni}, \citenamefont {Smart},\ and\ \citenamefont
  {Alavi}(2016)}]{doi:10.1021/acs.jctc.5b01190}%
  \BibitemOpen
  \bibfield  {author} {\bibinfo {author} {\bibfnamefont {G.}~\bibnamefont
  {Li~Manni}}, \bibinfo {author} {\bibfnamefont {S.~D.}\ \bibnamefont {Smart}},
  \ and\ \bibinfo {author} {\bibfnamefont {A.}~\bibnamefont {Alavi}},\ }\href
  {\doibase 10.1021/acs.jctc.5b01190} {\bibfield  {journal} {\bibinfo
  {journal} {J. Chem. Theory Comput.}\ }\textbf {\bibinfo {volume} {12}},\
  \bibinfo {pages} {1245} (\bibinfo {year} {2016})}\BibitemShut {NoStop}%
\bibitem [{\citenamefont {Li~Manni}\ and\ \citenamefont
  {Alavi}(2018)}]{doi:10.1021/acs.jpca.7b12710}%
  \BibitemOpen
  \bibfield  {author} {\bibinfo {author} {\bibfnamefont {G.}~\bibnamefont
  {Li~Manni}}\ and\ \bibinfo {author} {\bibfnamefont {A.}~\bibnamefont
  {Alavi}},\ }\href {\doibase 10.1021/acs.jpca.7b12710} {\bibfield  {journal}
  {\bibinfo  {journal} {J. Phys. Chem. A}\ }\textbf {\bibinfo {volume} {122}},\
  \bibinfo {pages} {4935} (\bibinfo {year} {2018})}\BibitemShut {NoStop}%
\bibitem [{\citenamefont {Li~Manni}\ \emph {et~al.}(2019)\citenamefont
  {Li~Manni}, \citenamefont {Kats}, \citenamefont {Tew},\ and\ \citenamefont
  {Alavi}}]{doi:10.1021/acs.jctc.8b01277}%
  \BibitemOpen
  \bibfield  {author} {\bibinfo {author} {\bibfnamefont {G.}~\bibnamefont
  {Li~Manni}}, \bibinfo {author} {\bibfnamefont {D.}~\bibnamefont {Kats}},
  \bibinfo {author} {\bibfnamefont {D.~P.}\ \bibnamefont {Tew}}, \ and\
  \bibinfo {author} {\bibfnamefont {A.}~\bibnamefont {Alavi}},\ }\href
  {\doibase 10.1021/acs.jctc.8b01277} {\bibfield  {journal} {\bibinfo
  {journal} {J. Chem. Theor. Comput.}\ }\textbf {\bibinfo {volume} {15}},\
  \bibinfo {pages} {1492} (\bibinfo {year} {2019})}\BibitemShut {NoStop}%
\bibitem [{\citenamefont {Weser}\ \emph {et~al.}(2020)\citenamefont {Weser},
  \citenamefont {Freitag}, \citenamefont {Guther}, \citenamefont {Alavi},\ and\
  \citenamefont {Manni}}]{Weser2020}%
  \BibitemOpen
  \bibfield  {author} {\bibinfo {author} {\bibfnamefont {O.}~\bibnamefont
  {Weser}}, \bibinfo {author} {\bibfnamefont {L.}~\bibnamefont {Freitag}},
  \bibinfo {author} {\bibfnamefont {K.}~\bibnamefont {Guther}}, \bibinfo
  {author} {\bibfnamefont {A.}~\bibnamefont {Alavi}}, \ and\ \bibinfo {author}
  {\bibfnamefont {G.~L.}\ \bibnamefont {Manni}},\ }\href {\doibase
  10.26434/chemrxiv.12411125.v1} {\  (\bibinfo {year} {2020}),\
  10.26434/chemrxiv.12411125.v1}\BibitemShut {NoStop}%
\bibitem [{\citenamefont {Vancoillie}\ \emph {et~al.}(2011)\citenamefont
  {Vancoillie}, \citenamefont {Zhao}, \citenamefont {Tran}, \citenamefont
  {Hendrickx},\ and\ \citenamefont {Pierloot}}]{doi:10.1021/ct200597h}%
  \BibitemOpen
  \bibfield  {author} {\bibinfo {author} {\bibfnamefont {S.}~\bibnamefont
  {Vancoillie}}, \bibinfo {author} {\bibfnamefont {H.}~\bibnamefont {Zhao}},
  \bibinfo {author} {\bibfnamefont {V.~T.}\ \bibnamefont {Tran}}, \bibinfo
  {author} {\bibfnamefont {M.~F.~A.}\ \bibnamefont {Hendrickx}}, \ and\
  \bibinfo {author} {\bibfnamefont {K.}~\bibnamefont {Pierloot}},\ }\href
  {\doibase 10.1021/ct200597h} {\bibfield  {journal} {\bibinfo  {journal} {J.
  Chem. Theor. Comput.}\ }\textbf {\bibinfo {volume} {7}},\ \bibinfo {pages}
  {3961} (\bibinfo {year} {2011})}\BibitemShut {NoStop}%
\bibitem [{\citenamefont {Radon}(2014)}]{doi:10.1021/ct500103h}%
  \BibitemOpen
  \bibfield  {author} {\bibinfo {author} {\bibfnamefont {M.}~\bibnamefont
  {Radon}},\ }\href {\doibase 10.1021/ct500103h} {\bibfield  {journal}
  {\bibinfo  {journal} {J. Chem. Theor. Comput.}\ }\textbf {\bibinfo {volume}
  {10}},\ \bibinfo {pages} {2306} (\bibinfo {year} {2014})}\BibitemShut
  {NoStop}%
\bibitem [{\citenamefont {Levine}\ \emph {et~al.}(2020)\citenamefont {Levine},
  \citenamefont {Hait}, \citenamefont {Tubman}, \citenamefont {Lehtola},
  \citenamefont {Whaley},\ and\ \citenamefont
  {Head-Gordon}}]{doi:10.1021/acs.jctc.9b01255}%
  \BibitemOpen
  \bibfield  {author} {\bibinfo {author} {\bibfnamefont {D.~S.}\ \bibnamefont
  {Levine}}, \bibinfo {author} {\bibfnamefont {D.}~\bibnamefont {Hait}},
  \bibinfo {author} {\bibfnamefont {N.~M.}\ \bibnamefont {Tubman}}, \bibinfo
  {author} {\bibfnamefont {S.}~\bibnamefont {Lehtola}}, \bibinfo {author}
  {\bibfnamefont {K.~B.}\ \bibnamefont {Whaley}}, \ and\ \bibinfo {author}
  {\bibfnamefont {M.}~\bibnamefont {Head-Gordon}},\ }\href {\doibase
  10.1021/acs.jctc.9b01255} {\bibfield  {journal} {\bibinfo  {journal} {Journal
  of Chemical Theory and Computation}\ }\textbf {\bibinfo {volume} {16}},\
  \bibinfo {pages} {2340} (\bibinfo {year} {2020})},\ \bibinfo {note} {pMID:
  32109055},\ \Eprint
  {http://arxiv.org/abs/https://doi.org/10.1021/acs.jctc.9b01255}
  {https://doi.org/10.1021/acs.jctc.9b01255} \BibitemShut {NoStop}%
\bibitem [{\citenamefont {Strauss}\ \emph {et~al.}(1985)\citenamefont
  {Strauss}, \citenamefont {Silver}, \citenamefont {Long}, \citenamefont
  {Thompson}, \citenamefont {Hudgens}, \citenamefont {Spartalian},\ and\
  \citenamefont {Ibers}}]{doi:10.1021/ja00300a021}%
  \BibitemOpen
  \bibfield  {author} {\bibinfo {author} {\bibfnamefont {S.~H.}\ \bibnamefont
  {Strauss}}, \bibinfo {author} {\bibfnamefont {M.~E.}\ \bibnamefont {Silver}},
  \bibinfo {author} {\bibfnamefont {K.~M.}\ \bibnamefont {Long}}, \bibinfo
  {author} {\bibfnamefont {R.~G.}\ \bibnamefont {Thompson}}, \bibinfo {author}
  {\bibfnamefont {R.~A.}\ \bibnamefont {Hudgens}}, \bibinfo {author}
  {\bibfnamefont {K.}~\bibnamefont {Spartalian}}, \ and\ \bibinfo {author}
  {\bibfnamefont {J.~A.}\ \bibnamefont {Ibers}},\ }\href {\doibase
  10.1021/ja00300a021} {\bibfield  {journal} {\bibinfo  {journal} {J. Am. Chem.
  Soc.}\ }\textbf {\bibinfo {volume} {107}},\ \bibinfo {pages} {4207} (\bibinfo
  {year} {1985})}\BibitemShut {NoStop}%
\bibitem [{\citenamefont {Sun}, \citenamefont {Yang},\ and\ \citenamefont
  {Chan}(2017)}]{SUN2017291}%
  \BibitemOpen
  \bibfield  {author} {\bibinfo {author} {\bibfnamefont {Q.}~\bibnamefont
  {Sun}}, \bibinfo {author} {\bibfnamefont {J.}~\bibnamefont {Yang}}, \ and\
  \bibinfo {author} {\bibfnamefont {G.~K.-L.}\ \bibnamefont {Chan}},\ }\href
  {\doibase https://doi.org/10.1016/j.cplett.2017.03.004} {\bibfield  {journal}
  {\bibinfo  {journal} {Chem. Phys. Lett.}\ }\textbf {\bibinfo {volume}
  {683}},\ \bibinfo {pages} {291 } (\bibinfo {year} {2017})}\BibitemShut
  {NoStop}%
\bibitem [{\citenamefont {Knizia}(2013)}]{Knizia2013}%
  \BibitemOpen
  \bibfield  {author} {\bibinfo {author} {\bibfnamefont {G.}~\bibnamefont
  {Knizia}},\ }\href {\doibase 10.1021/ct400687b} {\bibfield  {journal}
  {\bibinfo  {journal} {J. Chem. Theory Comput.}\ }\textbf {\bibinfo {volume}
  {9}},\ \bibinfo {pages} {4834} (\bibinfo {year} {2013})}\BibitemShut
  {NoStop}%
\bibitem [{\citenamefont {Knizia}\ and\ \citenamefont
  {Klein}(2015)}]{Knizia2015}%
  \BibitemOpen
  \bibfield  {author} {\bibinfo {author} {\bibfnamefont {G.}~\bibnamefont
  {Knizia}}\ and\ \bibinfo {author} {\bibfnamefont {J.~E. M.~N.}\ \bibnamefont
  {Klein}},\ }\href {\doibase 10.1002/anie.201410637} {\bibfield  {journal}
  {\bibinfo  {journal} {Angewandte Chemie Int. Ed.}\ }\textbf {\bibinfo
  {volume} {54}},\ \bibinfo {pages} {5518} (\bibinfo {year} {2015})},\ \Eprint
  {http://arxiv.org/abs/https://onlinelibrary.wiley.com/doi/pdf/10.1002/anie.201410637}
  {https://onlinelibrary.wiley.com/doi/pdf/10.1002/anie.201410637} \BibitemShut
  {NoStop}%
\bibitem [{\citenamefont {Booth}\ and\ \citenamefont
  {Chan}(2012)}]{doi:10.1063/1.4766327}%
  \BibitemOpen
  \bibfield  {author} {\bibinfo {author} {\bibfnamefont {G.~H.}\ \bibnamefont
  {Booth}}\ and\ \bibinfo {author} {\bibfnamefont {G.~K.-L.}\ \bibnamefont
  {Chan}},\ }\href {\doibase 10.1063/1.4766327} {\bibfield  {journal} {\bibinfo
   {journal} {J. Chem. Phys.}\ }\textbf {\bibinfo {volume} {137}},\ \bibinfo
  {pages} {191102} (\bibinfo {year} {2012})}\BibitemShut {NoStop}%
\end{thebibliography}
%\bibliographystyle{IEEEtran}
%merlin.mbs aipnum4-1.bst 2010-07-25 4.21a (PWD, AO, DPC) hacked
%Control: key (0)
%Control: author (8) initials jnrlst
%Control: editor formatted (1) identically to author
%Control: production of article title (-1) disabled
%Control: page (0) single
%Control: year (1) truncated
%Control: production of eprint (0) enabled
%

\end{document}